\documentclass{PoS}

\title{A review on Asteroseismology}

\ShortTitle{A review on Asteroseismology}

\author{\speaker{M. P. Di Mauro}\\
        INAF-IAPS Istituto di Astrofisica e Planetologia Spaziali, Roma, Italy\\
        E-mail: \email{maria.dimauro@inaf.it}}

\abstract{Over the last decade, thanks to the successful space missions launched to detect stellar pulsations,
 Asteroseismology has produced an extraordinary revolution in astrophysics, 
unveiling a wealth of results on structural properties of stars over a large
part of the H-R diagram.
 
Particularly impressive has been the development of Asteroseismology for
stars showing solar-like oscillations, which are excited and intrinsically damped
in stars with convective envelopes.

Here I will review on the modern era of Asteroseismology
with emphasis on results obtained for
solar-like stars and discuss its potential for the advancement of
stellar physics.
}

\FullConference{Frontier Research in Astrophysics ---  II\\
		23-28 May 2016\\
		Mondello (Palermo), Italy}

\begin{document}

\section{Introduction}
 Asteroseismology indicates the study of internal structure and dynamics of the stars from small oscillations observed at their surface. 
 The pulsations, visible as pattern of regions contracting and expanding on the photosphere,
 are produced by standing waves travelling inside the star which interfere constructively with themselves giving rise to resonant modes.
  These modes can be analyzed with the same mathematical techniques used in geophysics to probe
  the Earth's interior from the study of earthquakes.
  
  Each star can be thought as a musical instrument which plays a melody made
    of an original combination of modes. 
     The basic idea of Asteroseismology is to 
     recognize the size and shape of 'musical instruments' by recording the pitches or, in other words, by detecting
          the periodic motions of the stellar surface produced by the star-quakes.
   \begin{figure} 
              \centering 
               \includegraphics[width=4.7cm]{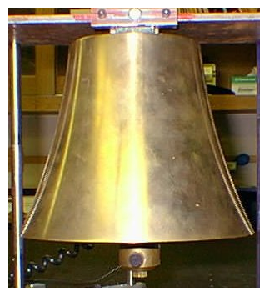}
               \includegraphics[width=4cm]{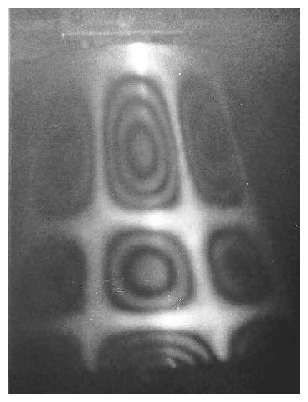} 
               \caption{Resonant acoustic modes of oscillations seen on the surface of a ringing bell by using holographic interferometry to see which parts vibrate in and out.}
               \label{bell}
               \end{figure}
         
\begin{figure} 
            \centering 
            \includegraphics[width=5.5cm]{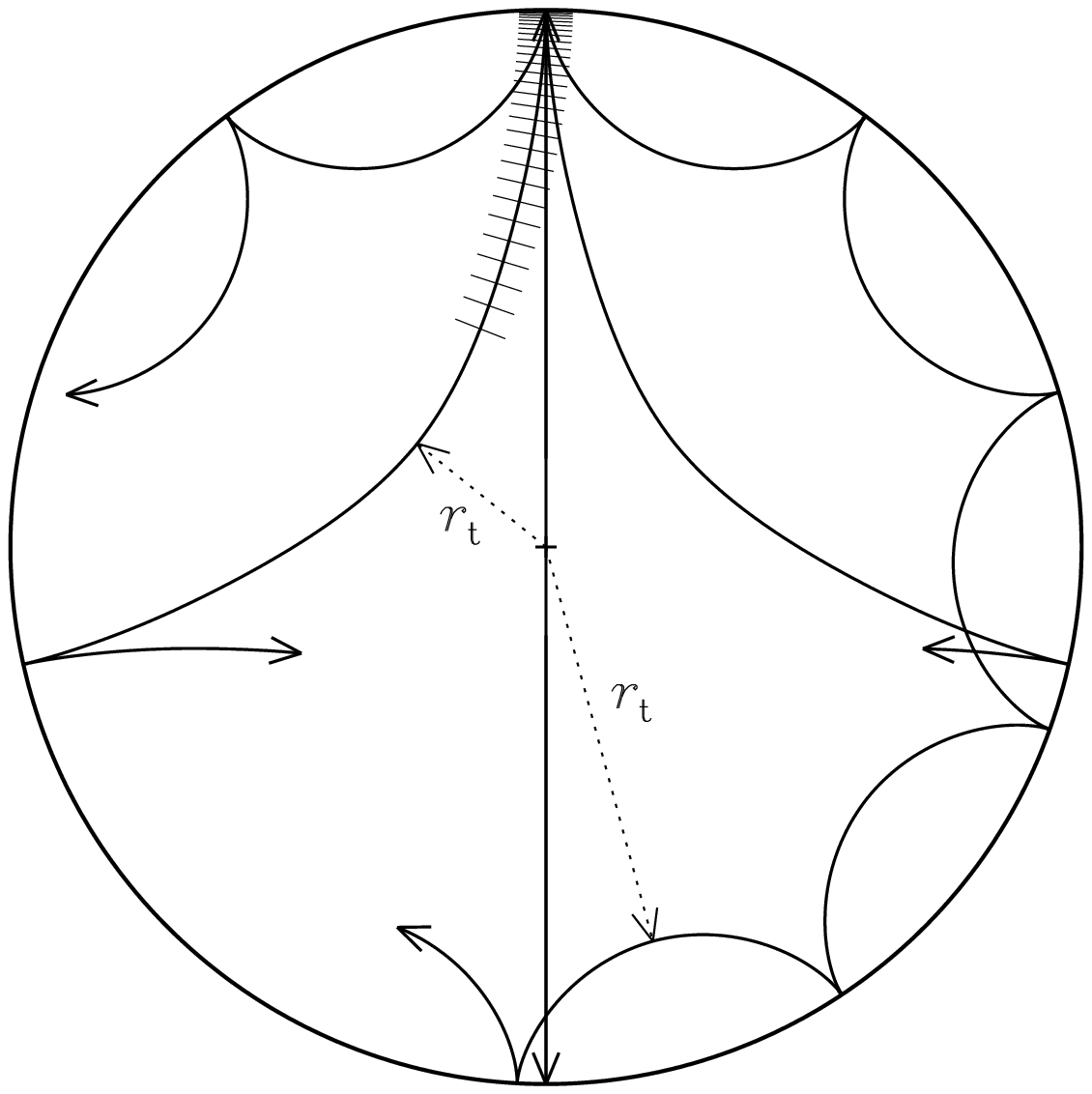}
            \includegraphics[width=3.6cm]{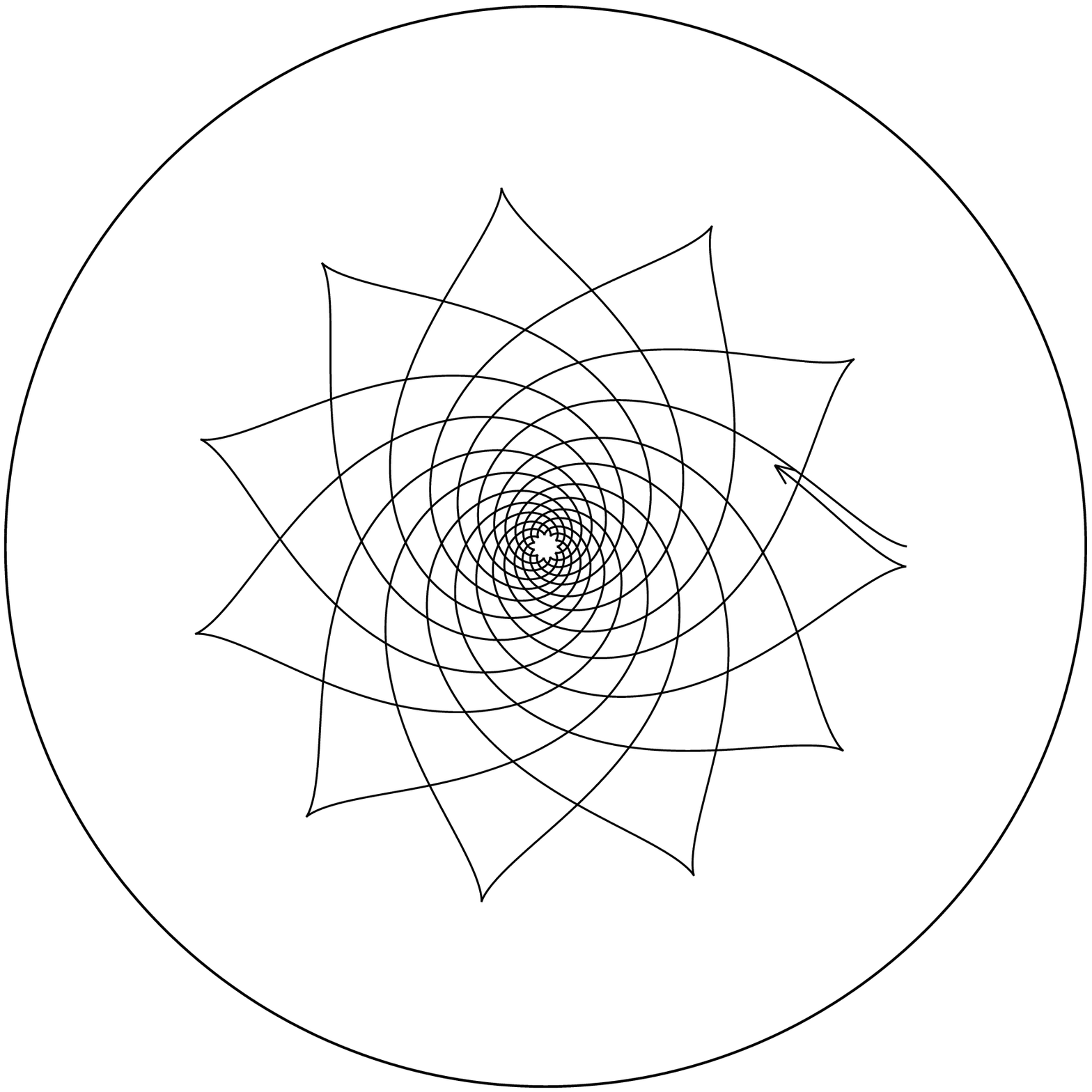} 
            \caption{Example of propagation of p modes (panel on the left) and g modes (panel on the right)  in the stellar interior.}
            \label{ray}
            \end{figure}

  The seismic waves supported in a star can be pressure or acoustic waves like in a bell (see Fig. \ref{bell}),
      but also internal or gravity waves like those occurring in a stratified medium such as the atmosphere or the ocean.
       They form the classes of 
     p and g modes respectively, named after the force that acts to restore the perturbed stellar equilibrium. Figure \ref{ray} shows a schematic view of acoustic and gravity waves propagating in the stellar interior.   
            
  The spatial configuration of the individual oscillation modes is defined by three numbers: 
  the radial
  order $n$ ($n=0,1,2,\ldots$), which is the number of 
  nodal surfaces in the radial direction,
  the harmonic degree $l$ ($l=0,1,2,\ldots$) and the azimuthal order $m$ ($m=-l,
  \ldots,l$), 
  which determine the behavior of the modes over
  the surface of the star in the directions of the latitude and of the longitude.
  
 Oscillations have several advantages over all the other observables:  their frequencies can be measured with high accuracy and depend in very simply way on the equilibrium structure of the stellar model; different modes propagate through different layers of the stellar interior.  
 Thus, a sufficiently rich spectrum of observed modes allows to probe the internal conditions at various depths inside a star and to test and revise models of stellar structure and theories of the evolution.

\begin{figure} 
\centering 
\includegraphics[width=8cm]{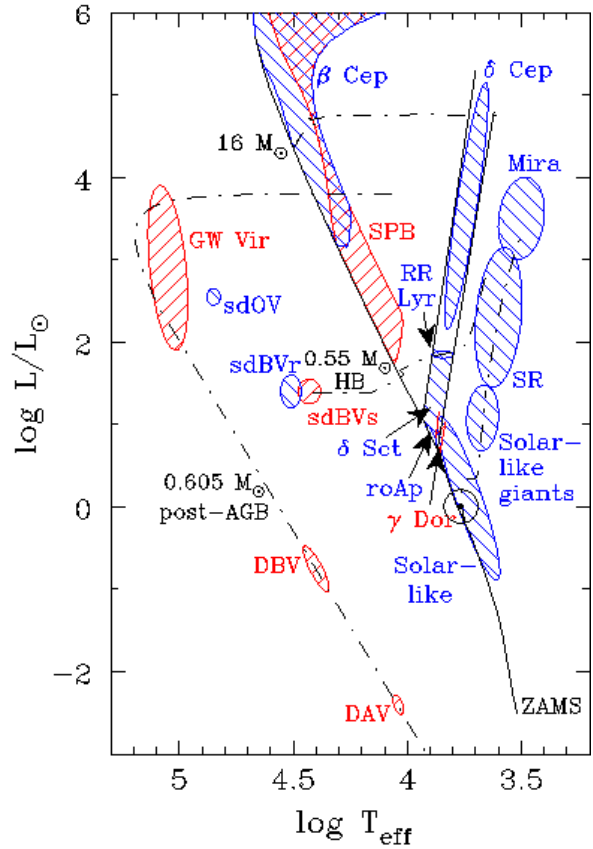} 
\caption{H-R diagram showing several selected classes of pulsating stars. Among them we can identify the classes of: solar-like stars, $\gamma$ Doradus, $\delta$-Scuti stars just on or little above the ZAMS; classical pulsators (RR Lyrae and $\delta$ Cephei), SPB and $\beta$ Cepheids along the upper part of the ZAMS; the evolved low mass hot subdwarfs and white-dwarf stars along the cooling sequence (DBV, DAV). Figure taken from \cite{Ha2012}.}
\label{HRall} 
\end{figure}

 Though the existence of pulsating stars and interpretation of their characteristics are well known since long time,
  the number of known classes of pulsating stars has recently highly increased.
Thanks to the ever improving precision in photometric and radial velocity measurements,
it has been possible to detect pulsational instability in stars of any evolutionary stage and spectral type from main-sequence to white dwarf cooling sequence (see Fig. \ref{HRall}).  

%\section{Driving mechanisms}
Stellar pulsations may be distinguished in self-excited and stochastic oscillations, according to the driving mechanism. 

The self-excited oscillations
arise from a perturbation to the energy flux resulting in 
a heat-engine mechanism converting thermal into mechanical energy.

If the valve which regulates the heat flux is provided by
variations of the opacity, the mechanism is named $\kappa$-mechanism and
 it depends on the existence of regions of partial ionization near the stellar surface.
 When the star shrinks, the energy of compression acts to raise the degree of ionization, the opacity increases and the gas heats up; during  expansion the opacity decreases and the heat is lost. This mechanism, firstly postulated by \cite{ba62} for explaining $\delta$ Cephei (also known as Cepheids)  variability, drives pulsations with intensity amplitude in the range of millimagnitudes in the classical pulsators, including white dwarfs,  $\delta$ Scuti stars, the rapidly rotating Ap stars, the $\beta$-Cephei stars, the slowly pulsating B (SPB) stars  and the
$\gamma$ Doradus. There is good evidence that this driving mechanism is also valid in other classes of stars, such as the Mira stars.

If perturbations to the energy flux depend on a strong temperature variation rising from thermonuclear reactions, then
the excitation mechanism is called  $\varepsilon$-mechanism.  Potentially, it could work in the stellar core of evolved massive stars, whose energy production is driven by the CNO cycle, but at present there isn't any observational evidence. 

The other major driving mechanism is the stochastic excitation, which works in the Sun and in the solar-like oscillators, where pulsations are excited and damped by
turbulent convection
and the oscillation modes are intrinsically stable. Although details of the stochastic excitation mechanism are not fully established, it is common belief that acoustic modes of few minutes period
  can be generated by convective cells which,
   in the shallow subsurface layer characterized by 
 superadiabatic stratification, can reach near-sonic velocity \cite{gol94}.
 An important aspect of the oscillations driven by
turbulent convection is that their excitation occurs at random times, and hence
the process is stochastic, with the effect that oscillations phase changes with time and the modes
 lifetime is finite.
 This is very
different from the heat engine mechanism, which instead excites pulsations coherently. This contrast is 
very useful for distinguishing between these two types of driving mechanisms in
the signal analysis. Another important attribute of stochastic driving is that all the possible resonant
modes typical of that star can be excited to observable amplitudes, while in the classical pulsators there exists 
 a selective mechanism which excites only some oscillation modes. 
 
\begin{figure} 
\centering 
\includegraphics[width=7cm]{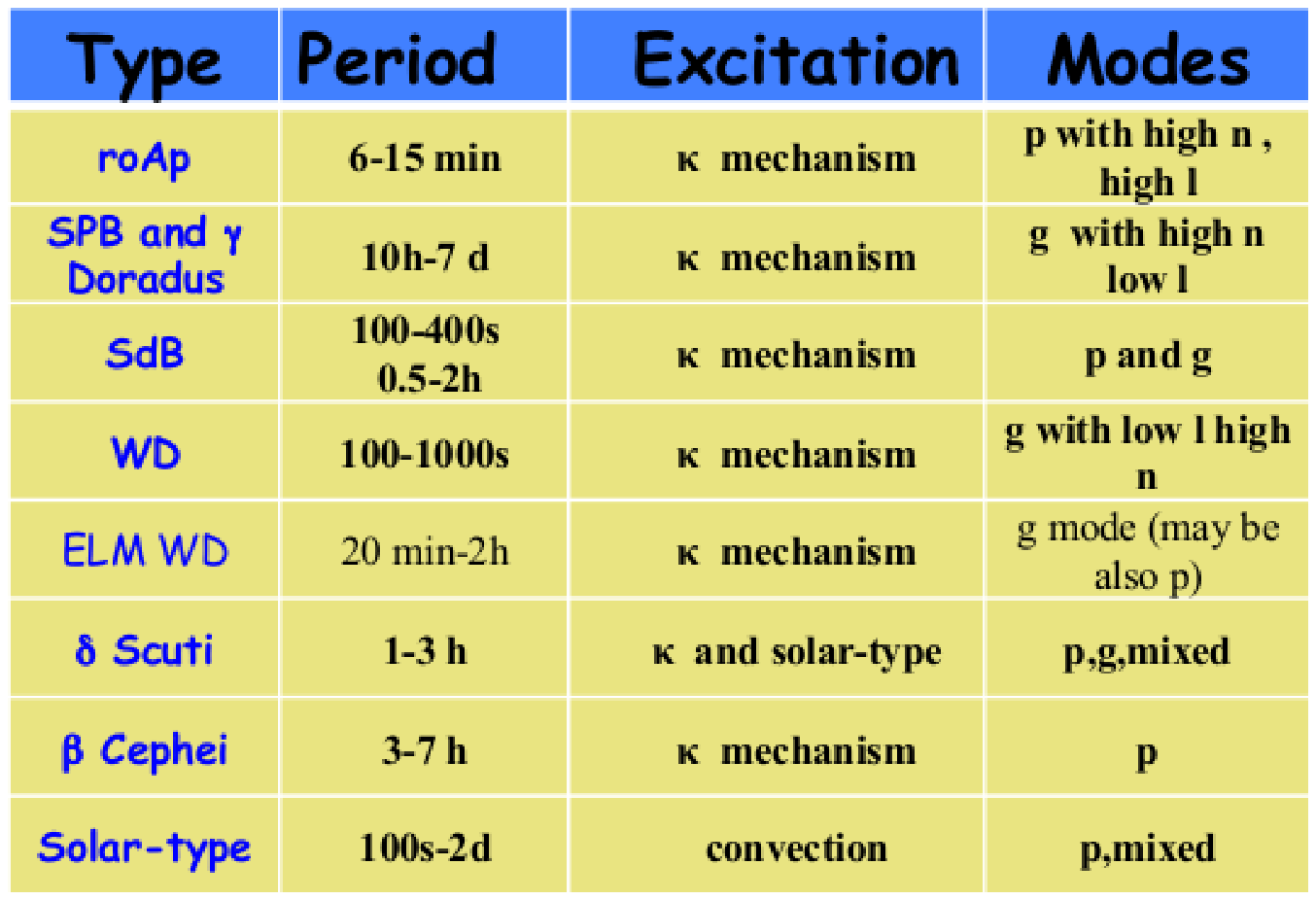} 
\caption{Main characteristics of selected classes of pulsating stars: stellar type , characteristic oscillations period, excitation mechanism, excited modes.}
\label{tab1}
\end{figure}

Main classification of known classes of pulsating stars are given in Fig. \ref{tab1} while a detailed description of the characteristics of the main
 asteroseismic targets across the H-R diagram can be found in, e.g., \cite{ae2010} or \cite{Ha2012}.
In addition, it is now largely recognized that some stars may show oscillations excited by two distinct mechanisms and hence can be considered "hybrid" pulsators. For instance, solar-like oscillations have been identified in the $\delta$-Scuti stars \cite{antoci}.

For basic concepts on the theory of stellar oscillations, the reader can 
 refer to classical books, e.g., \cite{Cox, Unno}; for a general review on asteroseismology, it is very useful the book by \cite{ae2010};
 for theoretical 
 methods in asteroseismology see the Lecture Notes on Stellar Oscillations by Christensen-Dalsgaard, the volume by \cite{Pijpers06} or
the review by, e.g.,  \cite{dima04}. 

The present paper provides 
a general overview on the recent observational and
theoretical successes obtained on the solar-type stars.
It should be pointed out, 
 that
 methods and techniques, developed and adopted for
solar-type stars, can be extended with success
 to other pulsating stars.
 
\section{Solar-like pulsators}

Stochastic oscillations, so-called solar-like oscillations, characterized by low amplitude (around $10 \,\mu$magnitudes or less), are mainly acoustic modes excited by turbulent convection like in the Sun, and
are predicted for all
 main-sequence and 
postmain-sequence stars cool enough to harbor an outer convective envelope.
Thus, solar-type stars are F, G and possibly K main-sequence and sub-giants stars. 
Moreover, solar-like oscillations have been found also in G, K and semi-regular M giants. Typical oscillation periods are of the order from few minutes in main-sequence stars, as in the Sun, up to about a few days in sub-giant and giant stars.

Although their presence in the spectrum of solar oscillations
has been debated for decades, 
theory predicts that, not only the p modes, but also the g modes can propagate in solar-like pulsators.
 Besides the remote possibility of detection due to the low amplitude of oscillations expected in atmosphere,
the major argument for years consisted
 in the difficulty in finding
and proving a possible mechanism for g modes excitation.
The controversy was closed 
in
 2005, when
 \cite{din} demonstrated  that excitation of internal gravity waves in cool stars with convective envelope is possible by the penetration of convective plumes into the
adjacent stably stratified radiative zone.
First hints of presence of gravity modes were reported after few years of operation of the instrument GOLF flying on board of the SOHO satellite by \cite{ga2002} and \cite{TC2004}. 
In 2007, after the analysis of
10 years of data collected by GOLF, a clear signature of g
 modes on the Sun was finally announced
by \cite{ga07, ga08}.  Recently, after two decades of full-disk measurements by GOLF and applying statistical techniques,
 Fossat et al. \cite{fossat2017} have succeeded 
not to observe g modes individually, but to provide a measurement of 
the period separation 
 and rotational splittings for very low-frequency g modes, in agreement with the predictions of the theoretical asymptotic approximations.
While these results on the Sun have been received with a certain scepticism by 
some part of the community, 
despite the difficulties in identifying g modes in main-sequence stars, 
 frequencies of internal gravity modes have been
  detected with no major problems 
 in stars more evolved, like the red giants (see Sec.~4.1)

\begin{figure} 
   \centering 
   \includegraphics[width=7cm]{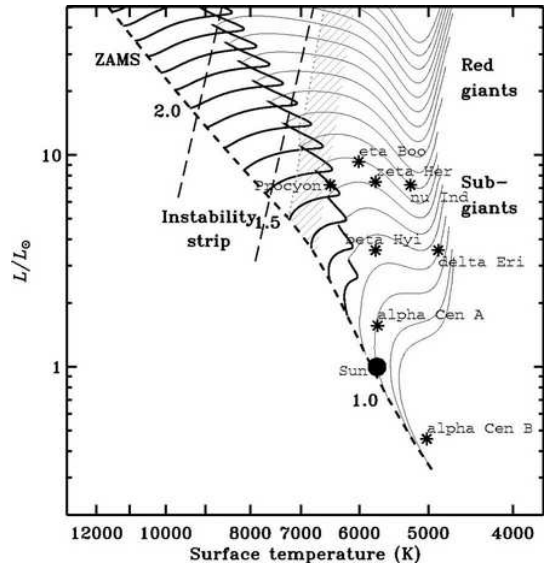} 
   \caption{H-R diagram showing location in luminosity and effective temperature of the solar-type pulsating stars. Asterisks indicate some targets on which solar-type oscillations have been already detected. The Sun is shown as a full circle. The solid lines are evolutionary tracks calculated for increasing values of mass. Figure taken from \cite{bedding03}.} 
   \label{HRsolartype} 
   \end{figure}

 Figure~\ref{HRsolartype} shows location in the
  H-R diagram of the solar-type pulsating stars with few typical solar-like targets.
 Some recent reviews  on asteroseismology of solar-like oscillators include those by \cite{bed11, CD11, mic12, cha13, dimauro2013}.

\section{Observations of solar-like oscillations}

The main asteroseismic observational data are the
 frequency and spatial configuration of the excited pulsation modes, which form the oscillation spectrum of a star.
   
The stellar oscillation spectra are extracted  
from time series acquired by
 two observational techniques: spectroscopy of velocity variations and photometry of stellar flux variations. These two techniques sample the physics of the oscillations differently: 
 radial velocity variations reveal the outward and inward
movement of the stellar surface due to stellar oscillations measured through the Doppler shift of
the spectrum lines; intensity variations reflect the brightness perturbations
of a star induced by stellar oscillations.

In order to identify the oscillation modes of a star, it is necessary to
collect time series characterized by high signal-to-noise ratio and high duty cycle,
because the analysis of the observations  and the identification of the oscillation modes can be strongly limited by presence of noise and gaps in the data. Moreover, the visibility of a mode depend directly on the relation between its lifetime and the total length and sampling time of the dataset. In fact, for their stochastic nature, solar-like oscillations have finite lifetime, being constantly damped and re-excited.
 
\begin{figure}
 \centering
 \includegraphics[width=10cm]{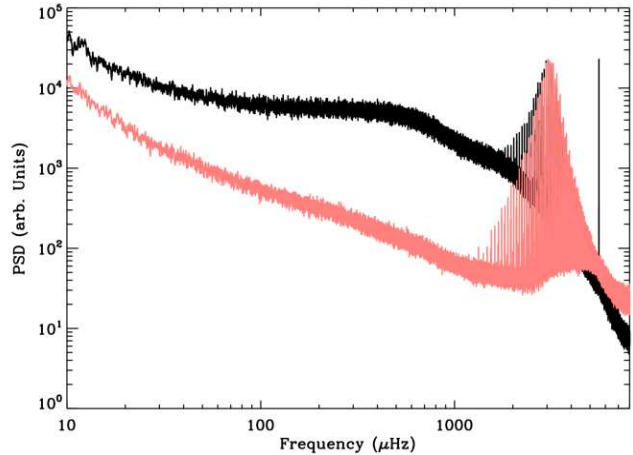}
  \caption{Comparison of the solar power spectrum taken from GOLF spectroscopic observations (in red)  
   and VIRGO photometry (in black),
    both instruments flying on-board of the SoHO satellite. Excess of power due to the oscillation is visible around $3000~\mu$Hz.  The solar convective background in the VIRGO data is clearly higher at lower frequencies
    compared to velocity measurement.
    Figure taken from \cite{garcia13}.}
    \label{backsun}
     \end{figure}

Spectroscopy, normally used for the Sun, provides better data, largely due to the fact that granulation on the surface of sun-like stars causes higher noise on photometric data than in spectroscopic data, as it is shown in Fig.~\ref{backsun}.
 Other stars may be different but, for the case of the Sun, Doppler measurements of spectrum lines formed at 200~km above the photosphere can reach a signal-to-noise ratio of about 300, which is ten times
  higher compared to intensity measurements obtained at disk center and with same spatial resolution \cite{ha88, garcia13}.

On the other hand since the observations are always the sum of all the pulsation modes simultaneously and
considering the stochastic nature of solar-like oscillations, which means that several modes, but not all are excited simultaneously, it becomes important the requirement of a long duty cycle, which is the need of continuous data sets for extended periods.
As a consequence, ground based spectroscopy is preferable for asteroseismology, however it is not the main source of the oscillation data.
In fact, ground based observations are subject to the day night cycle of the observing location or other periodic variations, such as seasonal visibility of stars or the non-uniform Earth's rotation around the Sun. Space based observations, generally with far superior duty cycles into respect to their ground based counterparts, allow to collect data for large numbers of stars and their 
instrumental requirements
 overlap with those of exoplanets finding missions.

Solar-type oscillations have been firstly found on the Sun in the early 60's, when Doppler velocity observations of the solar disk made by \cite{le62} showed 
 clear evidence of the presence of solar surface's oscillations with periods of about 5~min. 
Few years later,
more accurate observations carried out by \cite{de75} were able to confirm the theoretical hypothesis \cite{ul70, le71}
about the global
 character and the modal nature of the solar oscillations. This unprecedented discovery
   formed the basis for development of {\it Helioseismology}, based on the study of the solar oscillations spectrum: the combination of thousands of modes excited around the frequency of $3$~mHz, with a velocity amplitude of about $1~\mathrm{cm/s}$ and a brightness variation of about $8~ \mathrm{ppm}$.

During the last decades, Helioseismology has impressively changed our understanding of the structure, the internal dynamics and the temporal evolution of the solar interior. 
This progress has been possible thanks to the development of the interest in this discipline and the large quantity of observed pulsation modes detected by several
helioseismic experiments. In particular, high quality data 
 have been obtained since 1993 by 
  the IRIS  (International Research on the Interior of the Sun)  \cite{fossat}, the GONG (Global Oscillations Network Group) \cite{ha96} and the BiSON  (Birmingham Solar-Oscillations Network) \cite{chaplin96} networks,
consisting of a number of observing stations worldwide located at different latitudes, which allowed to monitor our star without temporal interruption. 

But the great success of Helioseismology arrived with the launch of the ESA/NASA SOHO spacecraft in 1996 and its three instruments: the Solar Oscillations Imager / Michelson Doppler Imager (SOI/MDI) \cite{sc95}, the Global Oscillations at Low Frequency (GOLF) \cite{gabriel5}  and the Variability of solar Irradiance and Gravity Oscillations (VIRGO) \cite{frolich}.  
The measurement of thousands of individual oscillations frequencies has
allowed to probe
the solar interior with high spatial resolution and establish 
that the predictions of standard solar model are remarkably close to the solar structure.
Helioseismology not only
 has allowed to improve the description of the relevant physics, such as equation of state, opacity table, nuclear reactions, but also to refine details like
 elements abundances, 
  relativistic effects, heavy-elements diffusion, overshooting, internal rotation, mixing.

Inferences
of the interior of stars other than the Sun 
appear to be much more complicated and less outstanding in terms of 
achievable results.
The large stellar distances, the point-source character of the stars and the very tiny amplitudes, make these oscillations hard to detect and restrict the asteroseismic studies
to the use of small sets of data often characterized by modes with only low harmonic 
degrees ($l\leq 3$).
% except in the cases of rapidly rotating objects.
Nevertheless, over the past decade,
Asteroseismology has developed as a consequence of the very important successes obtained by Helioseismology and thanks to
the improved quality of the seismic observations, from ground-based spectroscopy to the space missions launched in recent years and dedicated to the measure of stellar pulsations outside the atmosphere. 

The search for solar-like oscillations in stars has been ongoing since the early 80's. The first indication of solar-like oscillation power was found on $\alpha$Cmi (Procyon) by \cite{brown91}, while individual frequencies were firstly detected in 1995 on $\eta$Boo \cite{kjeldsen95}. Up to now solar-like oscillations have been observed from ground in numerous stars, despite the low signal-to-noise level in the data and the limitations in the available observing time at the large telescopes. Rarely it has been possible to organize observational campaigns in order to minimize the problems due to the gaps in the time series, however
it is not straightforward to combine data obtained by different instruments, as illustrated for the case of $\alpha$Cen A by \cite{bed04}.

The first dedicated Asteroseismology mission, launched successfully in the 2003 was MOST (Microvariability and Oscillations of Stars) \cite{walker03}, which
achieved great success with observations of classical pulsators. In the realm of solar-like oscillations, controversy was generated
when MOST failed to find evidence for oscillations in Procyon A
\cite{mat04}. 
However, the  results by
\cite{hub11} on Procyon~A, based on a simultaneous ground-based spectroscopic campaign \cite{are08} and high-precision photometry by the MOST
satellite \cite{gue08}, revealed
 that the problems rely in the modelling of the convective transport and the wrong estimation of the
oscillation amplitudes and mode lifetimes in stars somewhat more evolved than the Sun \cite{hou99}. MOST was intended to be a 1-year-long mission, but  it has produced about 5000 light curves and at present, after several years of operation, the Canadian space agency has just decided to switch it off waiting for financial resources.

The French-led CoRoT (Convection, Rotation \& planetary Transits) mission, launched in 2006, contributed substantially to boost this discipline
(e.g., \cite{app08}). 
 Before the launch of CoRoT, solar-like oscillations had been detected only in few main sequence stars. Analysis of CoRoT data disclosed solar-like oscillations, not only as expected in several low-mass main sequence stars, but also surprisingly in hundreds of red-giant stars and also in some massive main-sequence stars, like a 10~$M_{\odot}$ $\beta$- Cephei \cite{be09} and in a O-type star \cite{deg10}. Unfortunately, the instrument stopped to send data in October 2012 and on June 2013 the end of the mission was declared.

In March 2009, NASA launched the {\it Kepler} satellite, a
mission designed with the primary goal to search for
extra-solar planets \cite{borucki10} around distant Sun-like stars.
 During the four years of nominal operation, {\it Kepler}
 released photometry time-series data for about $\simeq 190,000$ stars 
enabling the asteroseismological study of several thousands of pulsating stars, see e.g. \cite{chaplin10}, including some exoplanet hosts, see e.g. \cite{CD10}. These data obtained for targets with spectral type from early F to late K represent a homogeneous set with unprecedent quality.
{\it Kepler} nominal mission ended in the 2013 with the breakdown of two of its four reaction wheels, but operations of {\it Kepler} are continuing as K2 mission, with the objective to observe, continuously for three-months periods, successive fields along the Ecliptic. The end of the K2 mission is expected sometimes before mid-2018 for fuel exhaustion.

Results obtained by the CoRoT and {\it Kepler} space missions will
be better discussed in details in the following sections.

\subsection{Analysis of observational data} 

The photometric and spectroscopic time series acquired during observations
of pulsating stars are used to generate frequency spectra. 
For a known function of time $f(t)$, the amplitude  as function of frequency $F(\omega)$ is obtained by applying Fourier-based methods.
The 'oscillation power spectrum' (see Fig. \ref{curvaluce})
 is given by the square of the amplitude in frequency.

If the signal-to-noise ratio is sufficiently high, the analysis of data in the frequency domain will reveal excited modes with amplitude above the
noise at the corresponding oscillating frequencies.

Time series containing multiple signals, like those due to solar-like oscillations, produce a rather complex Fourier spectrum. Moreover, the time series acquired are usually far from perfect, because the observations
 are not measured continuously and are sampled discretely.
  
  The gaps in the observed time series have to be avoided as best as possible, because they
  produce additional frequencies, so-called aliases, which contaminate the real oscillation signal. 
 
 The sampling rate should be chosen properly for each studied target, because it
 is directly related to the highest frequency which can be resolved in a power spectrum. 
 
 Moreover, the temporal length of the data set is crucial because
 the longer is the time series, the smaller is the frequency resolution. In the case of the Sun,
a period of at least one month is necessary to measure 
  the basic characteristics of its oscillation spectrum, including modes lifetime.
  
Thus,
the identification of which peaks correspond to true oscillation frequencies and which are sidelobes generated by the sampling
can be extremely hard and a number of methods have been developed to deal
with this problem. A more general discussion of several methods in use in  asteroseismology can be found in e.g., \cite{Pijpers06, chaba2017}.

\section{Properties of solar-like oscillations} 
\begin{figure}
\centering
\includegraphics[width=0.6\linewidth]{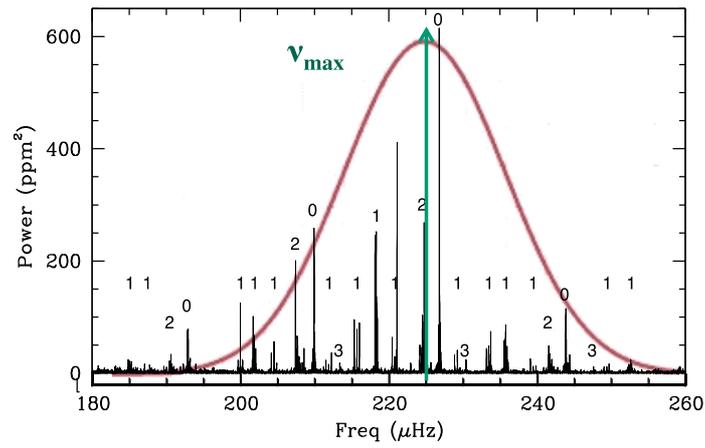}
\caption{Oscillation power spectrum with a typical Gaussian-like shape
for the red giant star KIC~4448777 observed by {\it Kepler}. The harmonic degrees of the modes ($l=0,1,2,3$) and $\nu_{\mathrm max}$ the frequency of maximum oscillation power are indicated.}
\label{curvaluce}
\end{figure}

\begin{figure}
\centering
\includegraphics[width=0.6\linewidth]{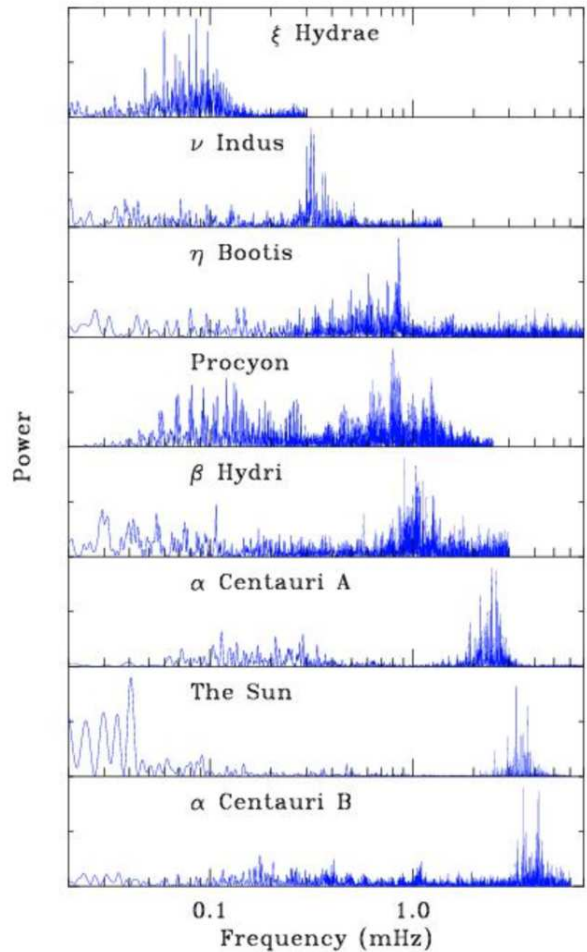}
\caption{Power spectra for different sun-like stars observed from ground based instruments.
The evolutionary state increases from the bottom to the top. Figure taken from \cite{bedding03}.}
\label{spettrisunlike}
\end{figure}

The observed oscillation power spectrum of solar-like stars is characterized by a typical Gaussian-like envelope, as shown in Fig. \ref{curvaluce}, and the frequency of maximum oscillation power is usually indicated by $\nu_{\mathrm max}$.
 From preliminary observations of solar-like oscillations on Procyon, Brown et al. \cite{brown91}
 conjectured that the frequency $\nu_{\mathrm max}$ could be related to the acoustic cutoff frequency $\nu_{ac}$, the highest frequency for acoustic modes,
 which defines the upper boundary of the p mode resonant cavities:
  \begin{equation}
  \nu_{\mathrm max}\propto \nu_{ac}\propto\frac{c}{H}\propto g T_{eff}^{-1/2}\propto \frac{M T_{eff}^{-1/2}}{R^2},
  \label{numax}
  \end{equation}
  where $c$ is the local speed of sound at radius $r$, $H=({\mathrm d} \ln \rho/ {\mathrm d}r)^ {-1}$ is the density scale height,  $T_{eff}$ is the effective temperature, $g$ is the surface gravity, $M$ is the stellar mass and $R$ is the photospheric stellar radius.
  According to Eq. \ref{numax}, the frequency $\nu_{max}$
    carries information on the physical conditions in the near-surface layers of the star.
Thus, as it has been well demonstrated both theoretically \cite{chaplin2008, belkacem2011} than observationally \cite{bedding03, stello08, bed11},
as a solar-type star evolves, its oscillation spectrum moves towards lower frequencies due to the decrease of the surface gravity (see Fig. \ref{spettrisunlike}).

Due to the point-like character of the sources,
oscillations modes, which can be observed in stars, are generally limited only to low harmonic degree.
In these conditions the Tassoul's asymptotic formulae \cite{tas80}, valid for $n\geq l$, is suitable for describing the properties of the stellar oscillations for p and g modes. 

\begin{figure} 
\centering
\includegraphics[width=0.8\linewidth]{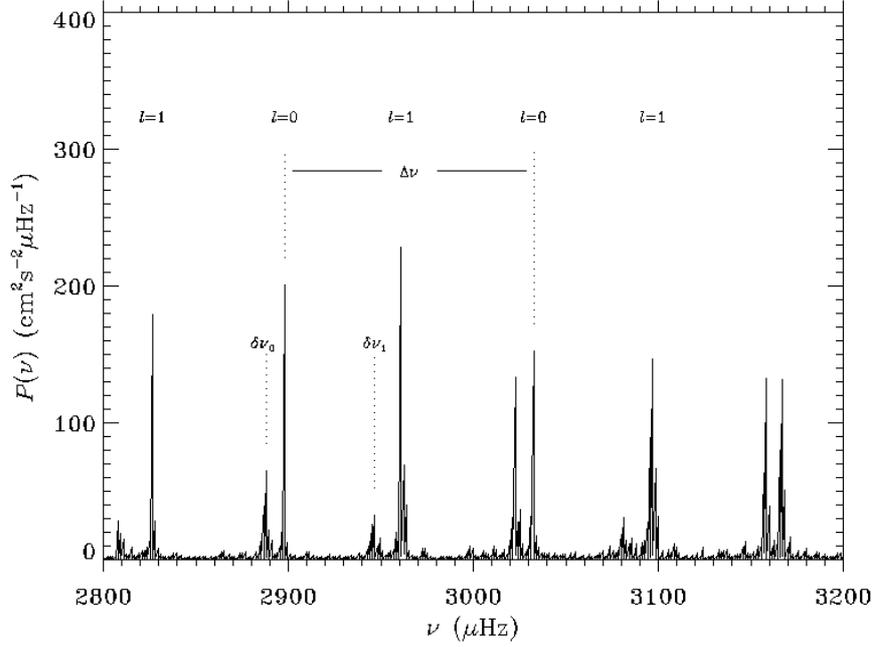} 
\caption{The oscillation power spectrum of the Sun obtained by BiSON. Figure taken from \cite{CD05}.} 
\label{spec} 
\end{figure}

Solar-like oscillations in main-sequence stars are generally p mode oscillations, characterized by high radial order and relatively high frequencies.  According to \cite{tas80},
the oscillation frequencies $\nu_{n,l}$ of an acoustic mode   
 of radial order $n$ and harmonic degree $l$   
should satisfy the relation:   
\begin{equation}   
\nu_{n,l}=\Delta\nu\left(n+\frac{l}{2}+\alpha+\frac{1}{4} \right)   
+\epsilon_{n,l} \; ,   
\label{eq1}   
\end{equation}   
where $\alpha$ is a function of the frequency determined by the
properties of the
surface layers, $\epsilon_{n,l}$   
 is a small correction which depends on the conditions in the stellar core.
$\Delta\nu$ is the inverse of    
the sound travel time across the stellar diameter: 
%***********************
\begin{equation}
\Delta\nu={\left(2\int_{0}^{R}\!\frac{{\rm d}r}{c}\right)}^{-1}. 
\label{Deltanu}
\end{equation}
%***********************
 To first approximation,
Eq. \ref{eq1} predicts that acoustic spectra should show
a series of equally spaced peaks between p modes of same degree,
whose frequency separation is
the so called large separation which is
approximately equivalent to
$\Delta\nu$:
%*************************
% EQUATION 3
%*************************
\begin{equation}
\Delta\nu \simeq\nu_{n+1,l}-\nu_{n,l}\equiv\Delta\nu_{l}\,.
\label{EQ_3}
\end{equation}
%*************************
As an example, the oscillation spectrum of the Sun is plotted in Fig. \ref{spec}.

In order to have an idea of how $\Delta\nu$
can easily be related to stellar properties, such as stellar mass $M$ and radius $R$, 
let us
suppose that the gas in stellar interiors can be described by
the perfect gas law and with
 fully ionized gas conditions, so that the adiabatic gradient  $\Gamma_1=\partial \ln
 p/\partial \ln
 \rho$ is constant and equal to $5/3$.
 Assuming the hydrostatic equilibrium, 
 with $G$ the gravitational constant,
it is easy  to demonstrate that: 
\begin{equation}
c=\frac{\Gamma_1 p}{\rho}\simeq\left(\Gamma_1\frac{\kappa_BT}{\mu m_u}\right)^{1/2}=
 (5GM/3R)^{1/2}
\end{equation}
where $\mu$ is the constant molecular weight,
$ \kappa_B$ the Boltzmann constant and $m_u$ the atomic weight.
 By substitution in the Eq. \ref{Deltanu}, then we obtain:
\begin{equation}
\Delta\nu=\frac{1}{2}\left(\frac{5}{3}\right)^{1/2}(GM)^{1/2}R^{-3/2}.
\label{large}
\end{equation}

The Eq.~\ref{large} shows that
$\Delta \nu$   
scales approximately as the square root of the mean density:
 as the star evolves the large separation
decreases with the increase of the radius.
%Also, the velocity amplitude increases with increasing $L/M$. 

In addition, the spectra are characterized by another series of peaks (see Fig. \ref{spec}), whose narrow separation is $\delta\nu_{l}$, known as small separation:
%*************************
% EQUATION 4
%*************************
\begin{equation}
\delta\nu_{l}\equiv \nu_{n,l}-\nu_{n-1,l+2}=(4l+6){\rm D}_{0}
\label{EQ_4}
\end{equation}
%*************************
where
%*************************
\begin{equation}
{\rm D}_{0}=-\frac{\Delta\nu}{4\pi^{2}\nu_{n,l}}\left[\frac{c(R)}{R}-
\int_{0}^{R}
\frac{{\rm d}c}{{\rm d}r}\frac{{\rm d}r}{r}\right] \, .
\end{equation}
%*************************
 The small frequency separation is sensitive to the sound-speed gradient in the core, which in turn is sensitive to the chemical composition gradient in   
 central regions of the star and hence to    
 its evolutionary state.

Concerning the internal gravity modes (g modes), their behavior is dominated by the buoyancy frequency
$N$: 
\begin{equation}
N^2=g\left(\frac{1}{\Gamma_1 p}\frac{{\mathrm d} p}{{\mathrm d} r}-\frac{1}{\rho}
\frac{{\mathrm d} \rho}{{\mathrm d} r}\right)\; ,
\end{equation}
where g is the local gravitational acceleration.
As for p modes, the relevant g modes are often of high radial order and according to
Tassoul's theory \cite{tas80}, in the asymptotic regime the g-modes of same harmonic degree are nearly uniformly spaced in period:
 \begin{equation}
\Delta P_{n,l}=\frac{N_0}{\sqrt{l(l+1)}}(n+\alpha_{l,g}),
\label{EQ_10}
\end{equation}
where
\begin{equation}
N_0=2\pi^2\left(\int_{r_1}^{r_2} N\frac{dr}{r}\right)^{-1},
\end{equation}
$r_1$ and $r_2$ define the region of propagation of the g modes and $\alpha_{l,g}$ is the phase term which depends on the details of the boundaries of the trapping region.

\begin{figure}
 \centering
\includegraphics[width=.8\linewidth]{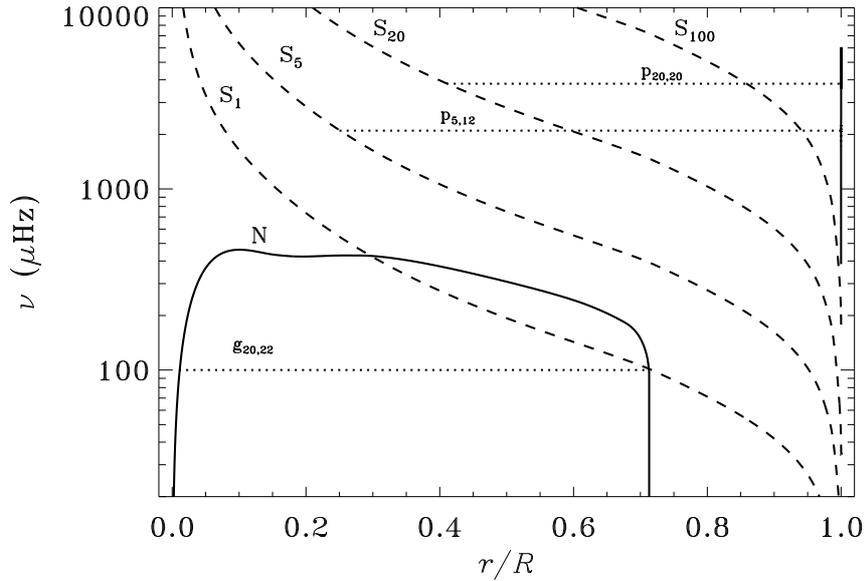}
\caption{Propagation diagram for p and g modes obtained for a standard solar model. The horizontal lines indicate the trapping regions  for a g mode with $\nu_{20,22}=100~\mu$Hz, and two p modes 
with $\nu_{5,12}=2097~\mu$Hz and $\nu_{20,20}=3808~\mu$Hz. The solid line indicates the buoyancy frequency $N$, the dashed lines indicate the characteristic acoustic frequencies for different harmonic degrees.}
\label{prop}
\end{figure}

The regions of propagation 
of p and g modes can be well illustrated by a propagation diagram, like in Fig. \ref{prop}. 
The characteristic acoustic frequency $S_l$ and
the acoustical cutoff frequency $\nu_c$  define respectively the lower and the upper limits of propagation of p modes with harmonic degree $l$.
The trapping region of g modes is delimited by the buoyancy frequency $N$, with $\nu^2_ {n,l}<N^2$.
In a main sequence star, like the Sun, acoustic modes propagate
at high frequencies through the convective zone up to the surface, while
gravity modes are trapped at low frequencies in the radiative
interior. In the convection zone $N^2<0$, hence outside the radiative region,
 the gravity waves 
 are evanescent, do not show oscillatory character in space and their amplitude decay exponentially.

\begin{figure}
\centering
\includegraphics[width=0.5\linewidth]{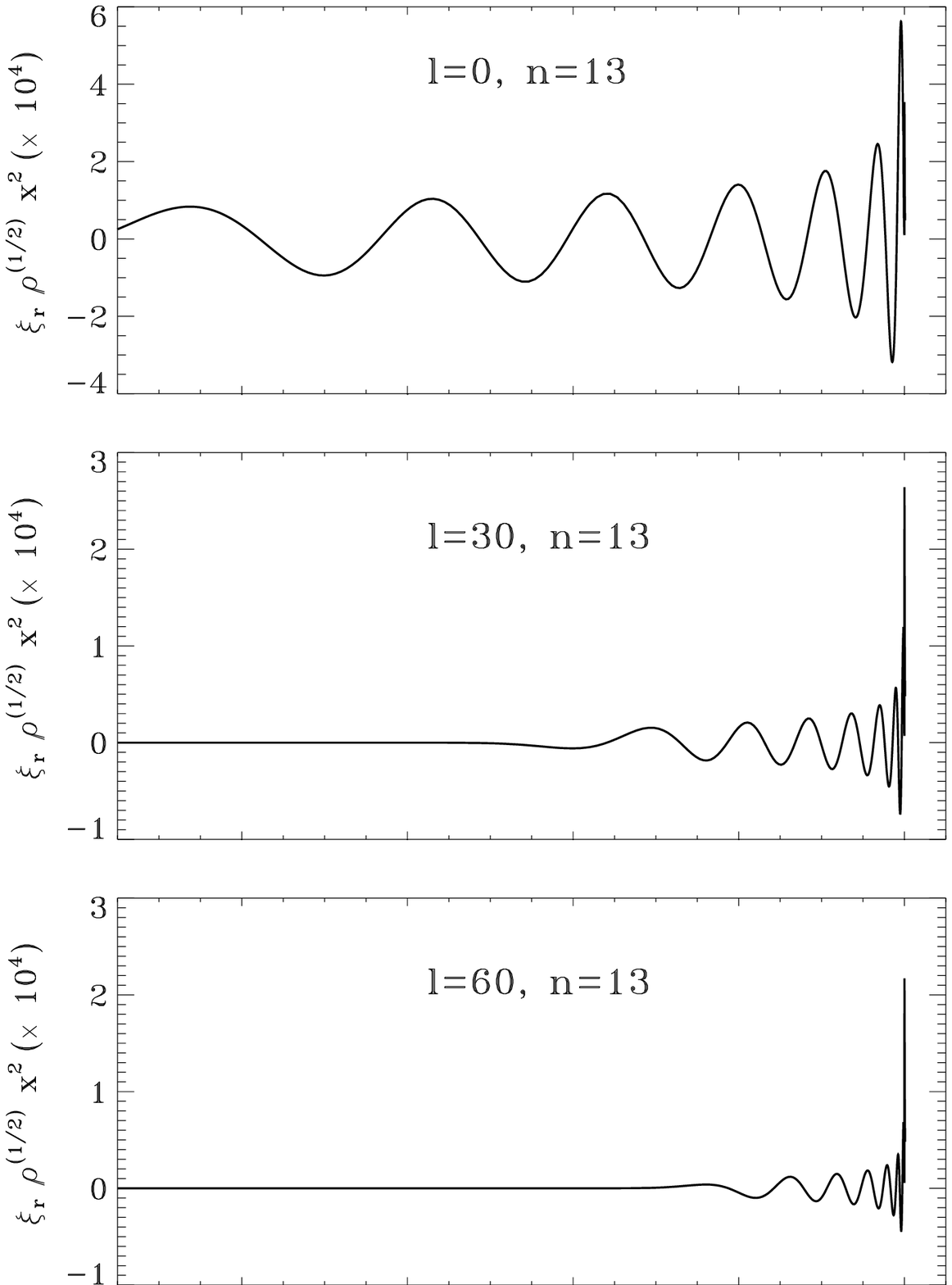}
\vspace{0.5cm}
\caption{Eigenfunctions for p modes with different harmonic degree
as function of the fractional radius $x=r/R_{\odot}$ 
in a standard solar model. Here, the oscillation behavior is enhanced by 
scaling the eigenfunctions with the square root of the density and the squared fractional radius.}
\label{aut}
\end{figure}

Figure \ref{aut} shows oscillation eigenfunctions for a selection of p modes with different harmonic,
degree calculated for a standard solar model.
The lower is the harmonic degree $l$,
the deeper is located the turning point of the acoustic mode.
Radial acoustic modes with $l=0$ penetrate to the centre, while the modes of
 highest harmonic degree are trapped in 
the outer layers.

\subsection{Mixed modes and seismology of evolved stars}
\begin{figure}   
\centering
\includegraphics[width=9cm]{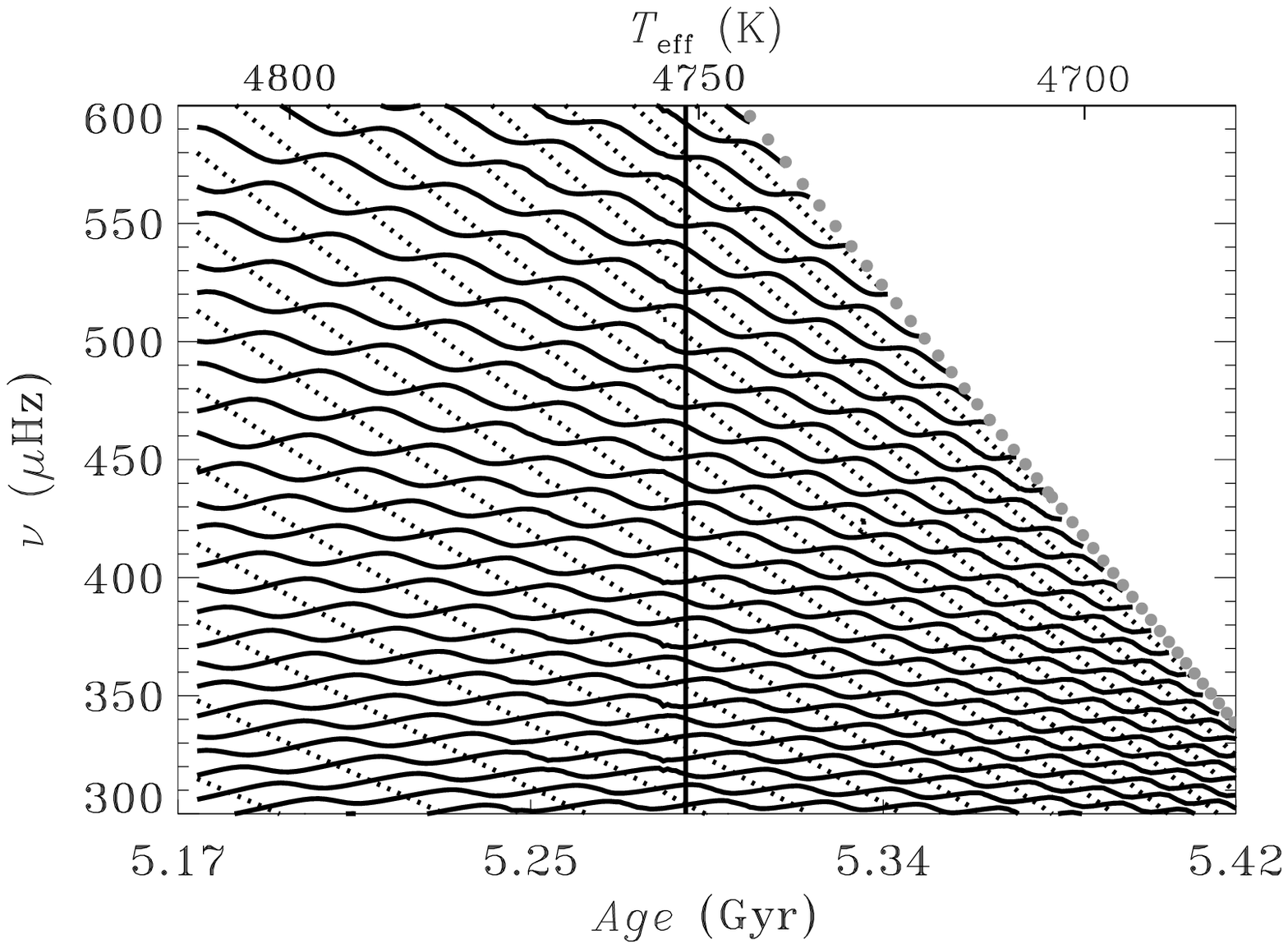}
\caption{Evolution of theoretical oscillation frequencies with age and with effective temperature of a stellar model computed with a mass
$M=1.32{\mathrm M}_{\odot}$. The dotted lines correspond to modes of    
degree $l=0$, and the solid lines to modes with $l=1$.   
 The grey dots indicate the acoustical cutoff frequency. The vertical line indicates the parameters of the observed red giant star KIC~4351319 observed by {\it Kepler} observed with {\it Kepler} \cite{dimauro2011}.}
\label{F.8}
\end{figure} 

\begin{figure}
\centering
 \includegraphics[height=7cm]{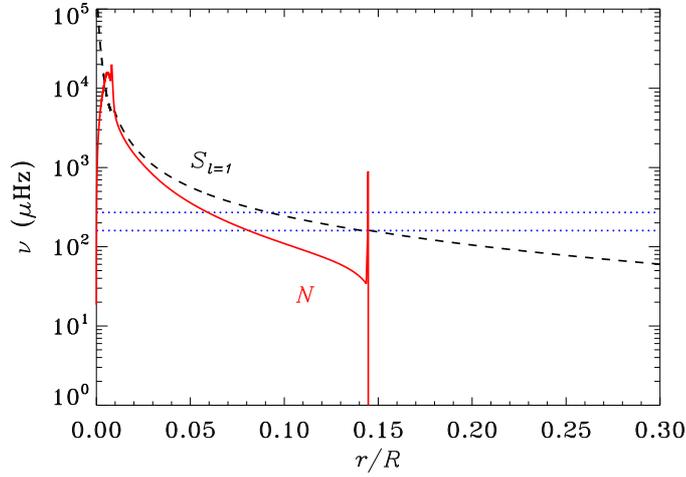}
\caption{Propagation diagram from the center to $ r=0.3\,R$ for a red giant model. The solid red line represents the buoyancy frequency $N$, constraining the region of propagation of g modes.
 The dashed black line represents the Lamb frequency $S_{l}$ for $l=1$, above which p modes are allowed to propagate in analogy to
 Fig.~10
  for main sequence stars. The dotted blue lines show the observed range of frequencies for the red giant star KIC~4448777 \cite{dimauro2016}.}
\label{propredgiant} 
\end{figure}
 
The properties of solar-like oscillations are expected to change
as the stellar structure evolves. 
According to Eq.~(\ref{eq1}) and considering that $\Delta\nu \propto R^{-3/2}$ from Eq.~(\ref{large}),
oscillation frequencies of a given harmonic degree should decrease as the star evolves and 
the radius increases   
and should appear almost uniformly spaced by $\Delta \nu$ at
each stage of evolution. However, in subgiants and red giants 
the frequencies of some non-radial modes appear to be shifted from the 
regular spacing due to the occurrence of the so-called `avoided crossing' \cite{dim03}, as it can be seen for the frequencies of $l=1$ in Fig. \ref{F.8}.

As the star evolves away from the main sequence, the core contracts 
and becomes denser 
causing a huge increase of the local gravitational 
acceleration and hence of
the buoyancy frequency in the deep interior of the star,
while the radius expands, causing a decrease of surface gravity and hence
 of the cut-off frequency.
As shown in the propagation diagram obtained for a red giant model in Fig. \ref{propredgiant}, the huge difference in density between the core region and the convective envelope, causes that all the trapped modes may be affected by the buoyancy frequency.
In these conditions, g modes 
propagate with high frequencies and can interact with p modes of similar frequency
and same harmonic degree, giving rise to modes with mixed character, 
which behave as g modes in the core and p modes in the outer envelope 
\cite{aizenman77}.
As a consequence, the spectrum of the red giants has a quite complicated appearance,
with a sequence of peaks uniformly spaced in frequency due to acoustic modes,
and other peaks with less clear pattern due to mixed modes.
Since the frequency of the modes increase as the harmonic degree increases,
the effect of 
the coupling becomes much weaker for modes with higher harmonic degree, due to the fact
that in these cases the gravity waves are better trapped in the 
stellar interior and hence better separated from the region of propagation 
of the acoustic waves \cite{dziembowski01}. 
\begin{figure}   
\centering  
\includegraphics[draft=false,scale=0.7]{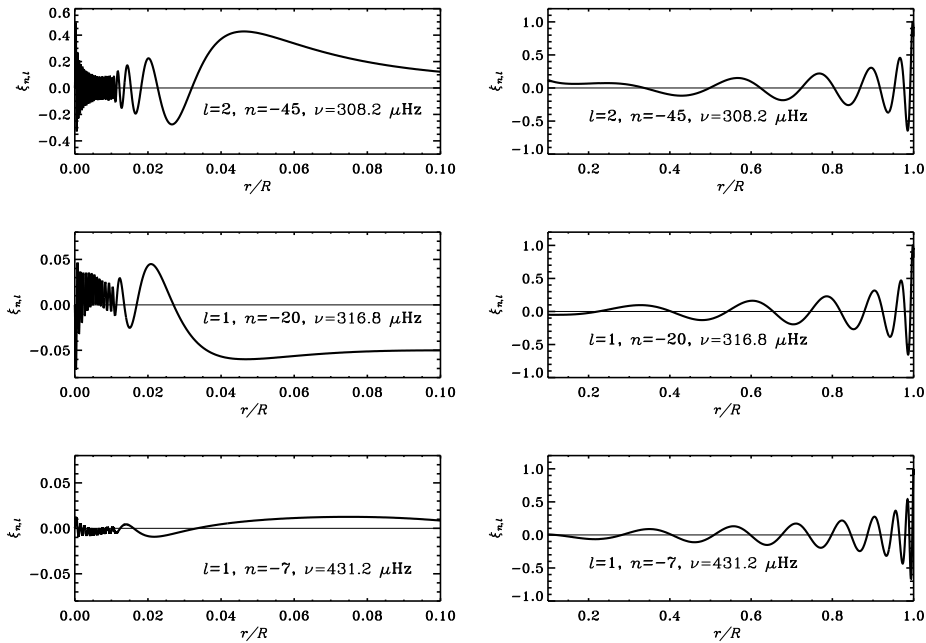}
\caption{Radial variations of eigenfunctions for three selected mixed modes calculated for a stellar model of a red giant star.
The left panels show the behavior of the eigenfunctions from the centre to $0.1\,R$. The right panels show the eigenfunctions from  $0.1\,R$ to the surface. The harmonic degree $l$, the radial order $n$ and 
the frequency  $\nu$ of the modes considered are indicated inside each panel.}
\label{eigen}   
\end{figure}   
Figure~\ref{eigen} shows the radial variation of the eigenfunctions of three mixed modes calculated for a structure model of a red giant star.
The upper panels of Fig.~\ref{eigen} show the eigenfunction for a mixed mode with high inertia and hence with a predominant gravity nature, characterized by a quite large amplitude in the core.
The middle panels of Fig.~\ref{eigen} show the eigenfunction for a mixed mode with both gravity and pressure character. 
The lower panels of Fig.~\ref{eigen} show the eigenfunction for a nearly pure acoustic mode, with an inertia comparable to that of the radial modes: the amplitude in the core is negligible.

It has been found by
\cite{montalban10} and observationally demonstrated by \cite{huber10}, that the scatter 
of $l=1$ modes caused
by `avoided crossing' decreases  as the star goes up to the red-giant branch:
as the luminosity increases and the core becomes denser, the $l=1$ acoustic
modes are better trapped and the oscillation spectra become more regular.
Once the star ignites He in the core, the core expands and the external 
convective zone becomes shallower which has the effect of increasing the probability of
coupling between g and p modes again.

 The presence of solar-like oscillations in red giants, firstly discovered by the space mission MOST \cite{barban07}, was well established by the CoRoT satellite
\cite{deridder09}, which was able to find
solar-like oscillations in a very large sample of G and K giant stars mainly lying in the core-helium-burning evolutionary phase ( e.g., \cite{hekker09}).  

The high-quality observations of the {\it Kepler} mission  enabled to detect solar-like oscillations in more than a thousand
red-giant stars from the red clump to the lower luminosity region of the
 red-giant branch, e.g. \cite{bedding10}, where stars are still burning H in the shell.
 
\subsection{Amplitudes of solar-like oscillations}

One of the greatest deficiencies in Asteroseismology of stars with 
surface convection zones is the lack of a proper theory 
to predict the expected amplitudes of oscillation.
This would be important to finally understand
convection and its interaction with pulsations and to outline the observational requirements.
 
Christensen-Dalsgaard \& Frandsen \cite{CDF83} made the first attempt to give 
a rough prediction of the solar-like oscillation amplitudes. 
They found velocity and luminosity amplitudes increasing with age and 
with mass along the main sequence, but
comparisons of prediction with observational 
measurements have always shown 
some inconsistency, indicating that 
there are still some contributions to damping so far ignored
 in the theory.
 Different attempts 
   made in the years to address this problem, led Kjeldsen \& Bedding \cite{kjeldsen95}
  to put forward a first empirical law based on observational data,
 by scaling amplitudes values from measurements of the Sun. 
They argued that the oscillation amplitude observed spectroscopically could  roughly be estimated as:
\begin{equation}
A_{vel}\propto\frac{L}{M},
\label{KB95}
\end{equation}
where $L$ is the stellar luminosity.
Tests of this law were initially biased by the fact that measurements of amplitudes had been obtained only for a limited number of solar-like stars.
The success of the {\it Kepler} mission,  with the detection of thousands of stars covering a large range of effective temperature and luminosity, enable astrophysicists to refine the relation (\ref{KB95}) to satisfy measurements for stars belonging to any evolutionary stage.
The inclusion of effects of granulation and mode lifetimes resulted in  different scaling laws of this type:
\begin{equation}
A_{vel}\propto\frac{L^s}{M^t}
\end{equation}
where $s$ and $t$ are two coefficients chosen in such a way   
 in order to estimate quite well observed amplitudes for field stars from main sequence to red clump with a precision of $25\%$ (see, e.g. \cite{huber2011}).

\section{Asteroseismic estimation of stellar properties}

\subsection{Radius and Mass by using main seismic parameters}
Preliminary asteroseismic studies are
 possible by using the main properties of the oscillation spectra, namely the large and small separations and the frequency of the maximum oscillation power, $\nu_{max}$. 
 This method is very useful to roughly estimate
 the global parameters of a star and for comparison of characteristics of
 a large sample of targets, because the main seismic properties can be extracted quite rapidly by using automated routines even from oscillation spectra with a low signal to noise ratio. 
 
A powerful and the most used seismic tool to roughly derive stellar mass and radius of stars is represented by the scaling laws
based
on the large
separation and the frequency of the maximum oscillation power $\nu_{max}$,
 like those
provided by \cite{kjeldsen95}:
\begin{equation}
\frac{R}{R_{\odot}}\simeq \Bigl(\frac{\nu_{max}}{\nu_{max \odot}}\Bigr)
\Bigl(\frac{\Delta\nu}{\Delta\nu_{\odot}}\Bigr)^{-2}
\Bigl(\frac{T_{eff}}{T_{eff\odot}}\Bigr)^{1/2}
\end{equation}
and

\begin{equation}
\frac{M}{M_{\odot}}\simeq
 \Bigl(\frac{\nu_{max}}{\nu_{max \odot}}\Bigr)^3
\Bigl(\frac{\Delta\nu}{\Delta\nu_{\odot}}\Bigr)^{-4}
\Bigl(\frac{T_{eff}}{T_{eff\odot}}\Bigr)^{3/2},
\end{equation}
where $\Delta\nu_{\odot}=134.9~\mu$Hz, $\nu_{max \odot}=3100~\mu$Hz and $T_{eff\odot}=5777~K$.
Other scaling laws, calibrated on a large sample of observed solar-like stars, have been obtained by  \cite{huber2011, mosser2013a}.
These tools allow to determine fundamental properties of the studied stars, such as mass and radius for main sequence stars with 5-7 \% of
uncertainty, 
%and age with an
%error up to 20\%, 
depending on the precision with which all the parameters are known, as demonstrated by \cite{miglio2012, huber2012}. 

The physical reasons for these scaling laws have not been definitely established, but there are different studies which aim to understand the physical relations among the quantities (see, e.g., \cite{huber2011, belkacem2011}) and in particular to test and improve the laws also for more evolved stars  by including other possible effects such as composition, angular momentum and probably also not well established chaotic effects.

The scaling laws are usually adopted
to carry out what is called 'Ensemble Asteroseismology' \cite{chaplin11}, a study of
similarities and differences in groups of few hundreds to thousands of stars, such as field stars or clusters.
Analysis of stars in open clusters is particularly interesting. In fact \cite{miglio2012} have been able to estimate the red-giant mass loss in two open clusters from determination of stellar masses in different evolutionary stages.

\subsection{Estimate of the age of a star}
Stellar ages cannot be determined by direct measurements and
 are very hard to estimate with high precision,
but this quantity is of great importance in astrophysics.
There are many methods to estimate
the age of a single star:  empirical
indicators such as stellar activity and gyrochronology which links rotation to age;
 photospheric lithium abundance; comparison of stellar model
isochrones with observed classical parameters.
However the
accuracy that can currently be reached by using all these methods is not satisfactory, not only because of the large errors in the estimates,
but also because better precision and accuracy can be reached by using seismic diagnostics \cite{garcia15}.

\begin{figure}
 \begin{center}
 \includegraphics[width=0.6\linewidth]{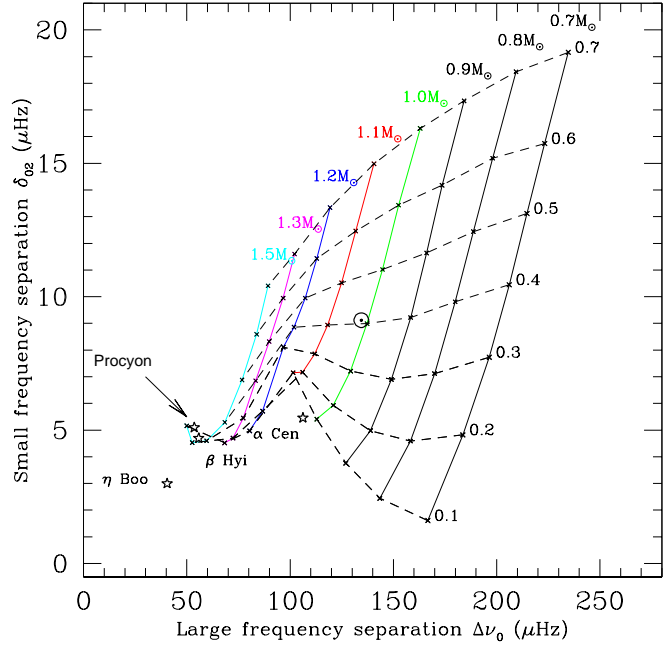} 
 \caption{C-D digram showing large versus small separations for a series of evolutionary
   models calculated for different masses, but fixed metallicity. Solid lines are evolutionary tracks
   for decreasing core's hydrogen abundance. Dashed black lines indicate models with same core's hydrogen content decreasing from $X_c=0.7$ (ZAMS) at the top, to $X_c=0.1$ at the bottom.  The
 Sun is marked by $\odot$. Black stars locate ground based observations of known solar-type stars.}  \label{CD} 
 \end{center}

 \end{figure}

Asteroseismic inferences on the age of
 main-sequence and post-main-sequence solar-type stars at a fixed composition can be obtained from the large and small frequency separations and by computing evolutionary
   tracks for fixed metallicity and mass deduced from the scaling relations.
This method allows to build the
  so-called 'C-D diagram' \cite{CD88}, like the one shown in Fig. \ref{CD}, similar to the H-R diagram showing evolutionary tracks of pulsating stars.
  Clearly the accuracy of this method depends on the accuracy with which the seismic parameters, effective temperature and metallicity are known for the specific target. 
  It was demonstrated by \cite{Chaplin2014} that this method, applied to
 a large sample of more than 500 solar-like core hydrogen
 burning stars, can provided relative age precisions of $10-15\%$.
 
 While the small frequency separation is an appropriate age
 indicator for the core hydrogen burning phase, this is no
 longer the case for evolved phases.
 In main-sequence stars, the tracks of large versus small separations for different masses and ages
are well detached (Fig. \ref{CD}).
For more evolved stars, the tracks converge and the small separation becomes
much less sensitive as a diagnostic, but there exist the possibility of a more powerful diagnostic. 
\cite{beck11} have demonstrated that the quality of {\it Kepler} observations gives the possibility to
measure the period spacings of mixed-modes with gravity-dominated character given by Eq. (\ref{EQ_10}),
which, like pure gravity modes, penetrate deeply in the core allowing
 to study the density contrast between the core region and the convective envelope and, like p modes, have amplitude at the surface high enough to be observed.
In particular, 
\cite{bedding11} found that measurements of the period spacings of the
 gravity-dominated mixed modes,
permit to distinguish stars with similar luminosity and effective temperature, but in different stages of evolution:
the hydrogen-burning stage and the helium burning phase.

A very important application of these tools for the red giants
is in Galactic archaeology, relating stellar age to the location of the stars in the Galaxy \cite{miglio2009, miglio2013, casagrande}. 
In fact, it is known that a variety of processes play an important role during the evolution: gas accretion, diffusion, migration of stars.  
Red giants represent a well-populated class of distance indicators, spanning a large age range, which can be used to study origin and chemical evolution of
 the Galactic disk in the regions probed  by CoRoT and {\it Kepler}.  
 
\subsection{Accurate stellar properties and internal details from individual oscillation frequencies}

More accurate and precise determination of fundamental parameters of a star can
be obtained by using observed individual pulsation frequencies, instead of the average oscillation properties of the spectra.

This method requires the detailed
computation of stellar models, in order to compare theoretical oscillation frequencies with the observed ones (see, e.g., \cite{mathur12, Metcalfe12}).
 Lebreton et al. \cite{lebreton} produced an extensive study in which quantified the impact of various assumptions in the input physics and in the free parameters for the calculation of
stellar models (e.g, mixing length, opacities, equation of state, helium abundance, initial mass etc..). This allowed them to estimate the
 precision which can be reached on stellar radius, mass and
  age of low-mass stars by using set of individual observed modes. 
Metcalfe et al. \cite{Metcalfe2014} performed
a similar, but far more extensive study of 42 solar-like
stars based on nine months of {\it Kepler} data. They found
that modelling based on  individual frequencies
typically doubles the precision compared to estimates based on the scaling
relations or on the large and small frequency separations.

\begin{figure}
\centering
\includegraphics[width=0.6\linewidth]{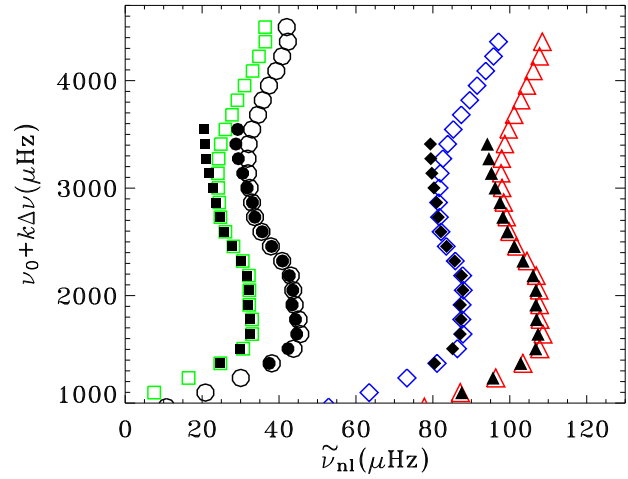}
\caption{\`Echelle diagram based on observed (filled 
symbols) and theoretical frequencies (open symbols) for the Sun. 
Circles are used for modes with $l=0$, triangles for $l=1$, squares for $l=2$, diamonds for $l=3$.}
\label{echesun}   
\end{figure}

\begin{figure}
\hspace{-0.5cm}
\centering
\includegraphics[width=0.6\linewidth]{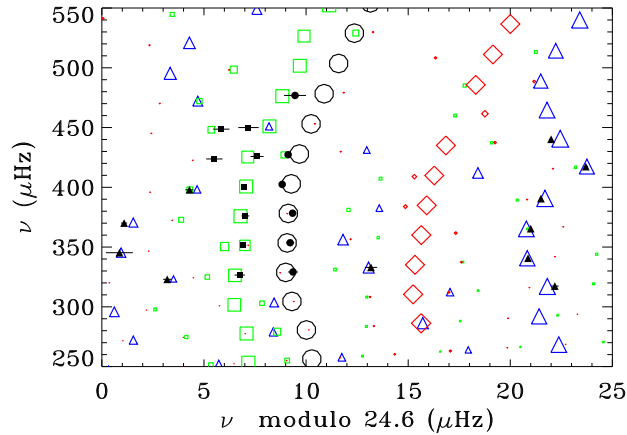}
\caption{\`Echelle diagram based on observed (filled 
symbols) and computed frequencies (open symbols) for the red-giant star KIC4351319 \cite{dimauro2011}. 
Circles are used for modes with $l=0$, triangles for $l=1$, squares for $l=2$, diamonds for $l=3$.      
The size of the open symbols indicates the relative surface amplitude of oscillation of the modes.}
\label{eche}   
\end{figure}

A detailed comparison between the theoretical 
and observed oscillation frequencies can be obtained by using the 
so-called {\it \'echelle diagram} \cite{grec83},
based on Eq. (1) and the asymptotic properties of the oscillation spectrum (see Fig. \ref{echesun} obtained for the Sun).
This diagram shows, 
for each harmonic degree, vertical ridges of frequencies in which consecutive symbols are equally spaced by the large separations. The distance between two adjacent columns of frequencies represents the small separation. 
Oscillation frequencies which do not follow the asymptotic relation, as in the case of gravity or mixed modes, show a significant departure from the regular pattern (see modes for $l=1$ in Fig. \ref{eche} obtained for a more evolved star).
The occurrence of mixed modes is a strong indicator of the evolutionary state of a star and the fitting of the observed modes with those calculated by theoretical models can provide, not only mass and radius with errors lower than $2\%$, but also an estimate of the age of a red-giant star with errors lower than $7\%$ \cite{dimauro2011}.

\begin{figure}
\vspace{-1cm}
\centering
  \includegraphics[height=11cm, angle=90]{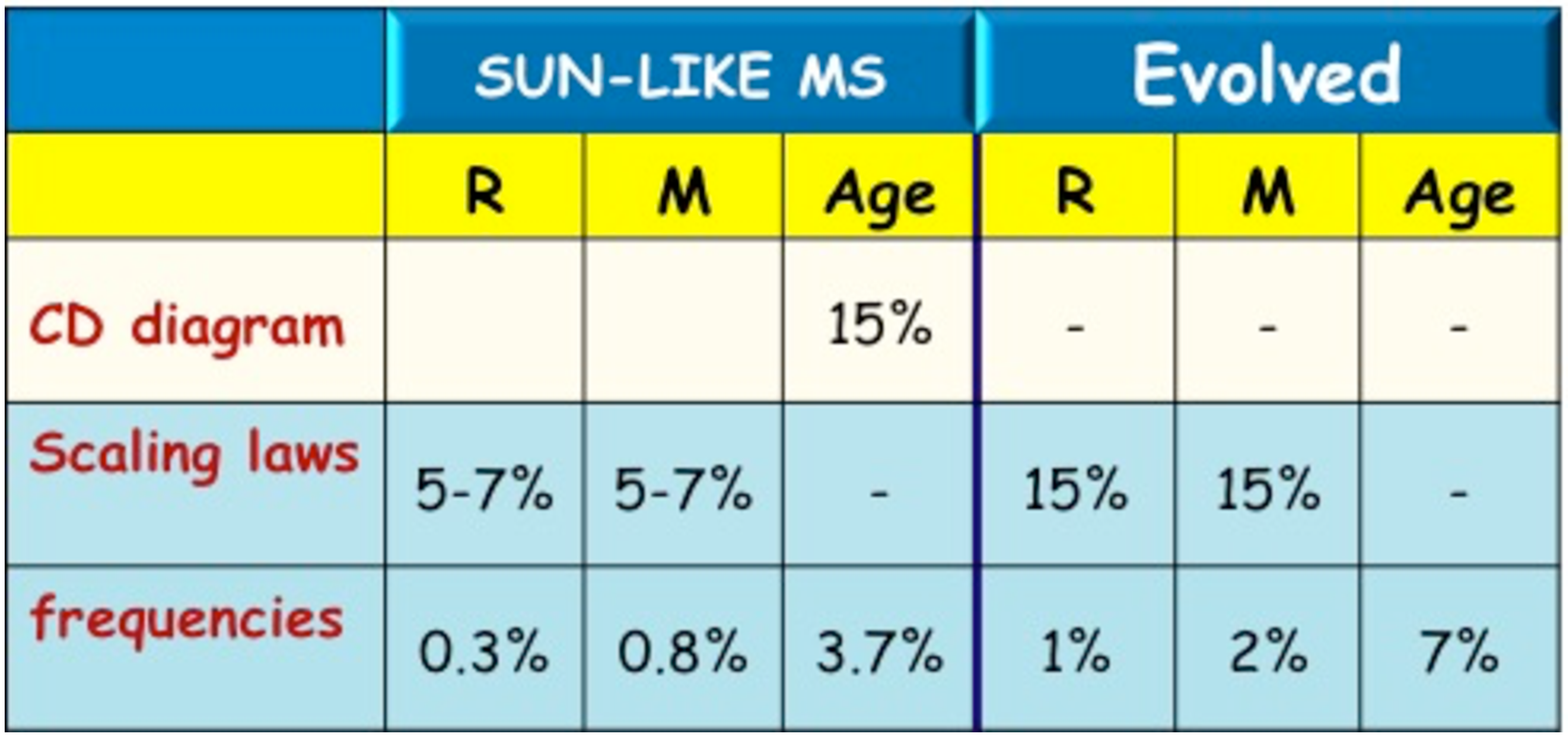}
  \vspace{-1cm}
\caption{Accuracy with which stellar parameters can be detected by using different seismic tools.
} 
\label{par}
\end{figure}

Figure \ref{par}
gives a schematic view on the accuracy with which stellar parameters can be measured by using asteroseismic tools for main-sequence and evolved stars.

In addition 
the \'echelle diagram shows, for each $l$, a weak oscillatory signal present in the observed and calculated p-modes frequencies which follow closely the asymptotic law, as shown in Fig. \ref{echesun} where symbols do not form straight columns.
The characteristics of such signal are related to the location and thermodynamic properties of glitches occurring inside the star.
Small periodic variations are produced, for example, by the borders of convective zones or by rapid changes in the first adiabatic exponent $\Gamma_1$, such as the one that occurs in the region of the second ionization of helium.
In main-sequence stars, signals
coming from different sharp features in the interior might
overlap, generating a complex behavior \cite{maz01}. 
 
Several attempts have been made in order to isolate the generated oscillatory components from the 
frequencies of oscillations or from a linear combinations of them and relate each signal to the specific sources allowing to determine, for example, the properties of the base of the convective envelope \cite{mon00, bal04} or the helium abundance in the stellar envelope \cite{lop97,mon98, per98, miglio03, basu04, houdek07}.

Application of this approach in solar-type stars, other than the Sun, were obtained for the first time by \cite{mazu2012}, who measured 
  the location of the base of the convective envelope
  and of the region of the second helium ionization in the main sequence star HD 49933  and by 
 \cite{miglio2010}, who extracted the He abundance in the red giant star 
 HR~7349, both stars having been observed by CoRoT.

\subsection{Characterization of exoplanets hosts}

The extreme photometric precision of the CoRoT and {\it Kepler} telescopes, suitable for accurate Asteroseismology, made these missions spectacularly successful also in their primary goal: the detection and characterization of extra-solar planetary systems by using the transit technique.  In particular, the {\it Kepler} mission, over its four years of nominal mission, monitored nearly 200,000 stars to detect transiting planets outside the Solar System (exoplanets) and especially Earth-size rocky planets. 

In the last few years the number of candidate planetary systems has increased continuously, also thanks to the contribution of ground-based high-precision spectroscopy performed by the twin spectrographs HARPS-S \cite{Major2003} and HARPS-N \cite{Cosentino2012} operating at La Silla (Chile) and La Palma (Spain), respectively.  
Thousands of planetary system candidates have been reported up to now, with nearly 700 of them confirmed and more than 3500 potential individual planets determined from the first three years of {\it Kepler} observations \cite{Batalha}.
 
Transit photometry and Doppler velocities, indirect methods for exoplanets detection, are able to provide properties of the planets, like for example the planetary radius and mass and the orbital radius, only as a function of the properties of the host star. To estimate mass and radius of an exoplanet and classify it as a gaseous giant body or as a telluric one, it is then necessary to estimate the mass and radius of the host-star, while to study the formation and history of an exoplanet it is necessary to measure the stellar age. 
Therefore, the accurate knowledge of the fundamental parameters of the host-stars is crucial for the interpretation of the detection of an exoplanet, the study of its structure and evolution and ultimately for its correct characterization as habitable planet.
 
The extreme efficiency of Asteroseismology in supporting the planetary search program for the accurate determination of the stellar fundamental parameters has been demonstrated by several recent works (e.g., \cite{van, borucki13, huber2013, Chaplin2013}).   
In addition, it has been proved  that spectroscopic radii for subgiants and giants are systematically lower by up a factor 1.5 into respect to the asteroseismic determination \cite{huber2013} .
On these bases, the asteroseismic revised stellar parameters lead to identify several tens of false-positive candidate planets, avoiding to waste precious resources for the ground-based follow-up, always necessary for the clear identification of an exoplanet.

\subsection{Internal rotation of stars}

The internal structure of a star at a given phase of its life is strongly affected by the angular momentum transport history. 
Unfortunately, physical processes that affect rotation and in turn are affected by rotation, such as convection, turbulent viscosity, meridional circulation, mixing of elements, internal gravity waves, dynamos and magnetism are at present not well understood and modeled with limited success \cite{marques2013, cantiello}.
Thus, investigating the internal rotational profile of a star and reconstructing its evolution with time become crucial in achieving basic constraints on the angular momentum transport mechanisms, acting in the stellar interior
during different phases of stellar evolution. 

\begin{figure}
 \vspace{-2.5cm}
\centering
  \includegraphics[width=0.7\linewidth]{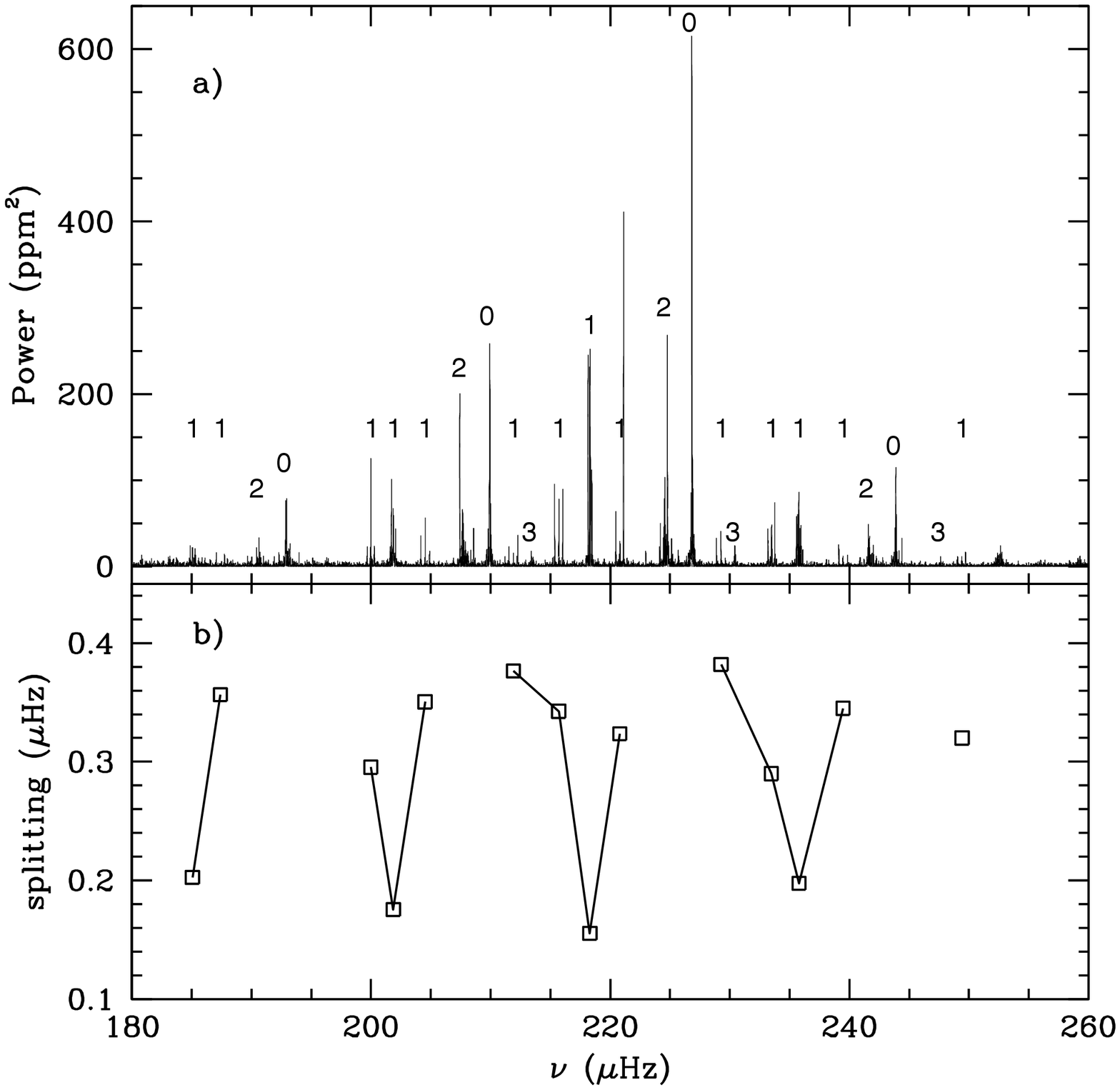}
  \vspace{-2.5cm}
\caption{The upper panel shows the oscillation spectrum of KIC~4448777 observed with {\it Kepler} from \cite{dimauro2016}. 
The harmonic degree of the observed modes
($l$=0,1,2,3) are indicated. Multiplets due to rotation are
visible for $l=1$. The lower panel shows the values of the
observed rotational splitting for individual $l=1$ modes.}
\label{fig:1}      
\end{figure}

Until fairly recently, rotation inside stars has been a largely unexplored field of research from an observational point of view. Over the past two decades Helioseismology changed this scenario, making it possible to measure the rotational profile in the Sun's interior through the measurement of the splittings of the oscillation frequencies. 
In a uniformly rotating star which oscillates with pulsation frequencies $\nu_{n,l}$, the rotation breaks the spherical symmetry of the stellar structure and
splits
the frequency of each oscillation mode of harmonic degree $l$ into 
$2l+1$ components, which appear as a multiplet in the power spectrum. Multiplets with a fixed radial order $n$ and harmonic degree $l$ are said to exhibit a frequency ``splitting'' defined by:
\begin{equation}
\delta \nu_{n,l,m}=\nu_{n,l,m}-\nu_{n,l,0}\; ,
\label{Eq.1}
\end{equation}
somewhat analogous to the Zeeman effect on the degenerate energy levels of an atom, where $m$ is the azimuthal order, as defined in Sec.~1.
Figure \ref{fig:1} shows
 the oscillation spectrum and the rotational splittings
for a red giant star studied recently by \cite{dimauro2016}.
By applying the standard perturbation theory 
 and under the hypothesis that the rotation of the star is sufficiently slow, so that effects of the centrifugal force can be neglected,  \cite{ledoux51} demonstrated that the Coriolis acceleration
 modifies the pulsations, so that the frequency separation between components of the multiplet is directly related to the angular velocity $\Omega(r,\theta)$ inside the star:
 %*********************
 \begin{equation}
 \Delta\nu_{n,l,m}=
m \Omega(r, \theta) (1-C_{n,l})
 \label{EQ_2}
 \end{equation}
 where $r$ is the radius, $\theta$
  is the colatitude and $C_{n,l}$ is the {\it Ledoux constant}, a structure parameter calculated on the non-rotating spherically symmetric model of the star. Thus, the dependence of the splittings on angular velocity  given by Eq. \ref{EQ_2} can be used to probe the stellar internal rotation.

The {\it Kepler} satellite, with photometric time series of unprecedented quality, cadence and duration has provided us with precious data for studying the internal rotational profile in a large sample of stars, characterized by a wide range of masses and evolutionary stages. 

In particular, evolved stars represent the ideal asteroseismic targets for probing the stellar internal rotation. In fact, red-giant frequency spectra, as seen in Sec. 4.1,
reveal mixed modes, 
which probe not only the outer layers, where they behave like acoustic modes, but also the deep radiative interior, where they
propagate as gravity waves.
Moreover, the red-giant phase represents a crucial step in the stellar angular momentum distribution history \cite{ceillier2013,marques2013}. When a star evolves off the relatively long and stable main sequence, its rotation starts evolving differently in the inner and outer parts causing the formation of a sharp rotation gradient in the intermediate regions, where hydrogen is still burning:  assuming that the angular momentum is locally conserved,
 the contraction of the core causes its rotation to speed up
  in a relatively short timescale, while the outer layers slow down due to their expansion. 
  Thus, the accurate determination of the rotational profiles in subgiants and red giants provides information on the angular momentum transport mechanism potentially leading to significant improvements in the
  modeling of stellar structure and evolution.
  
  High precision measurements of rotational splittings provided by {\it Kepler}  have shown that
the core in the red-giant stars
is rotating from a minimum of 8 to a maximum of 17 times 
faster than the upper layers  \cite{beck2012, beck2014, mosser2012c}.
 These results were confirmed by applying inversion techniques to rotational splittings by \cite{deheuvels2012, deheuvels2014, dimauro2016}.  Asteroseismology of large sample of stars   allowed to clarify that the mean core rotation 
 significantly slows down as stars ascend the red-giant branch \cite{mosser2012c, deheuvels2014}.

In addition,  Di Mauro et al. \cite{dimauro2016} demonstrated that the  rotational splittings can be employed to probe
 the variation with radius of  the angular velocity also within  the core, showing
 that in red giants
the entire core is rotating with a constant angular velocity and that an angular velocity shear layer is present in the region $0.007R\leq r \leq0.1R$, between the helium core and part of the hydrogen burning shell, in analogy to the tachocline discovered in the Sun. 
\begin{figure}
\centering
  \includegraphics[height=8cm]{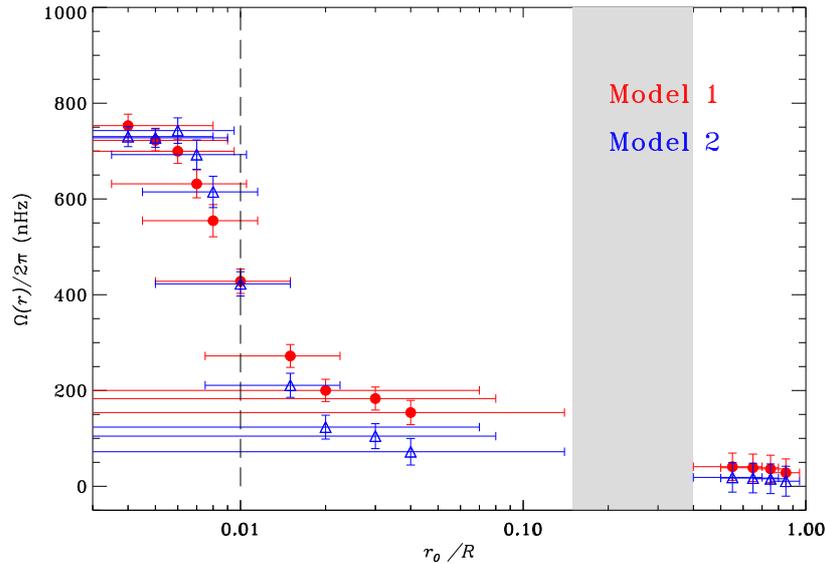}
\caption{Internal rotation of KIC~4448777 at different depths as obtained by inversion based on two best-fitting models.
Horizontal error bars indicate the radial resolution and depend on the region pervaded by the used set of data.
Vertical error bars show the uncertainty in the solutions and
are 2 standard deviations calculated on the observational errors.
 The dashed line indicates the
location of the inner edge of the H-burning shell.
The shaded area indicates the region inside the star in which it was not possible to determine any solution.
} 
\label{rot}
\end{figure}
The inferred rotation rate obtained by applying inversion techniques for  two models of the red giant star KIC~4448777 is shown
in Fig. \ref{rot}, where the points indicate the angular velocity against the selected target radii $\{r_0\}$.

All these results have shown that the internal rotation rates, predicted by current theoretical models of subgiants and red giants, are at least 10 times higher compared to seismic results, suggesting the need to investigate more efficient mechanisms of angular-momentum transport acting on the appropriate timescales during these phases of stellar evolution. Several theoretical investigations have explored the consequences of these results on internal angular momentum transport inside solar-like oscillating stars along their evolution. Different effects due to meridional circulation, shear instabilities, internal gravity waves and magnetic field have been explored, but found not sufficiently efficient  to extract enough internal angular momentum
 e.g., \cite{ceillier2013, marques2013, tayar2013, cantiello, fuller}. At the moment, only magneto-rotational instabilities of toroidal magnetic field, as demonstrated by \cite{rudiger2015}, have been found to be potentially efficient in the case of subgiant and early red giants.

\subsection{Stellar magnetic cycle}

It is well known that the p mode oscillations of the Sun vary with activity cycle providing diagnostic on the conditions below the photosphere. 
In analogy to the case of the Sun,
by analyzing the temporal evolution of oscillation parameters, mode frequency shifts and changes in the height of the p mode oscillations, it is possible to observe magnetic activity also in several solar-like stars.

 In particular, evidence of a
 stellar magnetic activity cycle taking place beneath the surface has been found in HD 49933 by \cite{garcia}. They revealed a modulation of at least 120 days, providing for the first time
 constraints for stellar dynamo models under conditions different from those of the Sun.

More recently \cite{salabert} investigating the photometric short cadence {\it Kepler} time series
and other spectroscopic observations found variability of about $1.5$~yr in the p mode frequencies of the young solar analog KIC 10644253, while \cite{kiefer} found signatures of stellar magnetic activity
 also in 6 solar-like stars over a sample of 24 studied targets, which were observed for at least 960 d each.

\subsection{Internal magnetic field in red giant stars}

One of the latest and controversial hit in this field has been the hypothetical disclosure of the presence of a 
 primordial internal magnetic field in some red giants by using only means of
Asteroseismology.
This possibility rises from observations of some red giants identified using {\it Kepler}'s photometry showing
a strange power spectrum characterized by 
 depressed dipole modes  \cite{stello2016}. These stars show normal radial modes (spherical harmonic degree $l= 0$),
but exhibit dipole ($l = 1$) modes with amplitude much lower than usual.
 In order to explain the suppression mechanism, Fuller et al. \cite{fuller} have shown that
 depressed dipole stellar
 oscillation modes can arise from a sort of magnetic greenhouse effect.
 The idea is that if a magnetic field is present in
 the core,  since the field cannot
 be spherically symmetric, the waves are scattered to high angular degree and become trapped
 within the core where they eventually dissipate.
 The authors found that magnetic fields should be stronger than $\simeq10^5~G$ in order to produce the observed depression.
  
  This idea, although very fascinating is still very much debated.
  Mosser et al. \cite{mos2017}
   proved that depressed dipole modes in red giants are just highly damped mixed modes and invalidated the hypothesis that depressed dipole modes result from the suppression of the oscillations in the radiative core of the stars due to strong internal magnetic fields.
   In fact, except for the amplitude of dipolar modes, seismic properties of the stars with depressed modes are equivalent to those of normal stars. 
   
\section{Conclusions}

The recent development of Asteroseismology, driven by new satellites observations of unprecedented quality and scope, 
has put the Sun into a broader context, 
 confirming that 
techniques and tools developed for Helioseismology
can be applied with success to other stars and showing that a real
synergy exists between our star and the others.

 The outstanding results provided by Asteroseismology have clearly demonstrated the potential of stellar pulsations to unveil the hidden interior and to study fundamental parameters, such as mass, radius and age of main sequence and also more evolved stars. 
Among the several striking findings which have risen great interest in the stellar physics community, we should mention:
 the proof 
that internal gravity waves are excited in the interior of solar-like stars and that they can be better detected in evolved targets
than in main-sequence stars; secondly, that red-giant stars 
 can be probed into details, as contrary to the main-sequence stars and the Sun itself, thanks to the use of the mixed modes which can be measured at surface. 
Furthermore, having proved to be able to measure 
the core's rotation in evolved stars, 
it appears not far the moment in which it will be possible to understand
how stellar evolution modifies rotational properties and
the angular momentum is conserved 
 as
a star advances towards the helium-core burning phase.

Asteroseismology provides, without doubts, an extensive possibility for testing and understanding the physical processes occurring in a star and there is high expectation from the 
data that will be obtained in the next future for stars 
with different pulsational characteristics.
From the observational point of view,  the situation is expected to significantly improve in the future
with the realization of the ground-based network SONG
(Stellar Observations Network Group) \cite{Grundahl2006},
which will deliver precise multi-site radial-velocity time series
for asteroseismology and exoplanet studies of nearby stars. At present only one of the four planned fully robotic telescopes is already operating, having collected between the period 2014-2015 excellent data on the subgiant $\mu$~Herculis, while a second telescope, located in western China, is in commissioning phase.

From the space, there is great expectation for the launch in June 2018 of TESS (Transiting Exoplanet Survey Satellite) \cite{ricker2014}, a NASA space mission with the objective to scan the whole sky over a two-year period to search for exoplanets and carry out Asteroseismology. It will observe several fields for about 1 month consecutively, while observations of the ecliptic poles will last 1 year.  
Finally, the ESA satellite PLATO 2.0  (PLAnetary
Transits and Oscillation of stars)  \cite{rauer2014}, whose launch is planned for the 2024, will detect and characterize terrestrial planets around solar-like stars, providing stellar radii and masses with an accuracy of $2\%$ and ages with an accuracy $10\%$.  The PLATO 2.0 instrument consists of 34 small aperture telescopes providing a wide field-of-view much larger than {\it Kepler} and a large photometric magnitude range (4-16 mag).
The major advantage of both TESS and PLATO 2.0  into respect of {\it Kepler} is the fact that these missions will focus on  nearby bright stars.

It is evident that the resulting improvements in stellar characterization and modelling will be indeed crucial for broad areas of astrophysics, including the investigation of the structure and evolution of the Galaxy and the understanding of the formation of elements in the Universe. Several theoretical knots are still waiting to be better understood and explained, such as stellar convective motion and magnetic dynamo action.
\section*{Acknowledgments}
 I dedicate this manuscript to my beloved father and to my supervisor Prof. Lucio Patern\'o,
who both sadly passed away very recently while I was writing this review.

\bigskip
\bigskip
\noindent {\bf DISCUSSION}

\bigskip
\noindent {\bf W. Becker:} 
What is the typical damping time? What is the damping mechanism for acoustic modes?

\bigskip
\noindent {\bf MARIA PIA DI MAURO:} Damping time for solar-like stars is typical of the order of few minutes or hours, however
 damping mechanism is poorly understood. So far, it seems that damping is dominated by 
  effects of convection and the exchange of energy between the pulsations and turbulent convective motions.
Some magnetic cycle effect might also be important, but  
 discrepancy between theory
and observation indicates that most likely something in the theory is still missing. For more details see review by
\cite{hou15}.

\bigskip
\noindent {\bf Dimitri Basilavo:} 
\noindent {\bf :} Does the hot jupiter exoplanet influence on the registered asteroseismological signals?

\bigskip
\noindent {\bf MARIA PIA DI MAURO:} Hot Jupiter, or close stellar companions in binary system,
 can cause tidal effects which produce second order effects on small oscillations. The magnitude of the effects, often negligible, depends on 
 orbit's parameters.
 
 \bigskip 
\noindent {\bf Anonymous speaker:} 
 \noindent {\bf :} What kind of creative treatment of the {\it Kepler} data did the {\it Kepler} team perform since they disclosed nearly 2000 exoplanets in one go?
 
 \bigskip
 \noindent {\bf MARIA PIA DI MAURO:} It takes long time to confirm an exoplanet candidate because it requires spectroscopic observations from the ground. The required time for follow up is typically of one year during which several targets are observed. Besides this, the {\it Kepler} team has used some statistical calculation to give an estimation of the exoplanets present in the observational data.

\end{document}